\documentclass{article}

\usepackage{arxiv}
\usepackage{comment}
\usepackage[export]{adjustbox}
\usepackage[utf8]{inputenc} 
\usepackage[T1]{fontenc}    
\usepackage{hyperref}       
\usepackage{url}            
\usepackage{booktabs}       
\usepackage{amsfonts}       
\usepackage{nicefrac}       
\usepackage{microtype}      
\usepackage{lipsum}		
\usepackage{graphicx}
\usepackage{natbib}
\usepackage{doi}
\usepackage{xcolor}

\title{DNN-ForwardTesting: A New Trading Strategy Validation using Statistical Timeseries Analysis and Deep Neural Networks}

\author{ \href{https://orcid.org/0000-0002-3843-386X}{\includegraphics[scale=0.06]{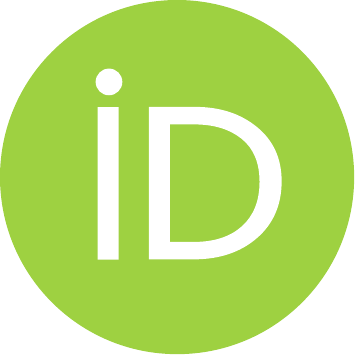}\hspace{1mm}Ivan Letteri}\thanks{\url{https://www.ivanletteri.it} } \\
	Department of Information Engineering, Computer Science and Maths\\
	University of L'Aquila\\
	L'Aquila, Italy, 67100 \\
	\texttt{ivan.letteri@univaq.it} \\
	\And
	\href{https://orcid.org/0000-0003-2327-9393}{\includegraphics[scale=0.06]{orcid.pdf}\hspace{1mm}Giuseppe Della Penna} \\
	Department of Information Engineering, Computer Science and Maths\\
	University of L'Aquila\\
	L'Aquila, Italy, 67100 \\
	\texttt{giuseppe.dellapenna@univaq.it} \\
	\And
    \href{https://orcid.org/0000-0001-9521-4711}{\includegraphics[scale=0.06]{orcid.pdf}\hspace{1mm}Giovanni De Gasperis}\\
	Department of Information Engineering, Computer Science and Maths\\
	University of L'Aquila\\
	L'Aquila, Italy, 67100 \\
	\texttt{giovanni.degasperis@univaq.it} \\
	\And
    \href{https://orcid.org/0000-0003-0329-2419}{\includegraphics[scale=0.06]{orcid.pdf}\hspace{1mm}Abeer Dyoub}\\
	Department of Information Engineering, Computer Science and Maths\\
	University of L'Aquila\\
	L'Aquila, Italy, 67100 \\
	\texttt{abeer.dyoub@univaq.it} \\
}

\date{}

\renewcommand{\undertitle}{}
\renewcommand{\shorttitle}{\textit{arXiv} Template}

\hypersetup{
pdftitle={A template for the arxiv style},
pdfsubject={ai-algo trading.NC, ai-algo trading.QM},
pdfauthor={Ivan Letteri et al.},
pdfkeywords={First keyword, Second keyword, More},
}

\begin{document}
\maketitle

\begin{abstract}

In general, traders test their trading strategies by applying them on the historical market data ( \textit{backtesting}), and then apply to the future trades the strategy that achieved the maximum profit on such past data.

In this paper, we propose a new trading strategy, called \textit{DNN-forwardtesting}, that determines the strategy to apply by testing it on the possible future predicted by a deep neural network that has been designed to perform stock price forecasts and trained with the market historical data.

In order to generate such an historical dataset, we first perform an exploratory data analysis on a set of ten securities and, in particular, analize their volatility through a novel k-means-based procedure. 
Then, we restrict the dataset to a small number of assets with the same volatility coefficient and use such data to train a deep feed-forward neural network that forecasts the prices for the next 30 days of open stocks market.
Finally, our trading system calculates the most effective technical indicator by applying it to the DNNs predictions and uses such indicator to guide its trades.

The results confirm that neural networks outperform classical statistical techniques when performing such forecasts, and their predictions allow to select a trading strategy that, when applied to the real future, increases Expectancy, Sharpe, Sortino, and Calmar ratios with respect to the strategy selected through traditional backtesting.
\end{abstract}


\tableofcontents

\section{Introduction}
\label{sect:intro}
Stock market forecasting is considered a research field with promising returns for traders and investors. However, there are considerable challenges to predicting stock market trends accurately and precisely enough due to their complexity, chaotic and non-linear nature. Indeed, traditional statistical models, which have been extensively applied to the market trend prediction so far, can easily handle only linear or stationary sequences.

According to the Efficient Market Hypothesis, stock prices reflect all past information, so the share prices themselves and the trading volume are sufficient to predict future price movements. Indeed, since new information is unpredictable and stock prices are adjusted immediately after the information is made public, the stock price follows a \textit{random walk} \cite{Eugene1965Fama} or, more generally, a stochastic process). Therefore, we first used traditional time series modelling techniques such as the Autoregressive Integrated Moving Average (ARIMA) model to predict stock prices. However, even if ARIMA model has its strength in robustness and efficiency in terms of \textit{short-term} forecasting (\cite{Adebiyi2014Ayodele}), most ARIMA models are univariate, so they usually model only the historical stock price series (\cite{LUX20071808}). On the other hand, in our work, we aim at an exhaustive prediction by means of price action trading activity (also called ``\textit{naked trading}''), i.e., considering daily information on maximum, minimum, opening, and closing prices and trading volume. 

Artificial intelligence methods are being currently employed in a variety of tasks, for example to classify cyber attacks (e.g., \cite{LetteriPG18}, \cite{Gonzalo2021Caasas}), predict network traffic anomalies (e.g., \cite{Letteri2019journal}, \cite{Letteri2019journal}), and predict the course of a disease. Among such AI methods, artificial neural networks (ANN) and, in particular deep neural networks (DNN) proved suitable for dealing with complex nonlinear problems with multiple influencing factors (\cite{Letteri2022lod21})). Indeed, they are often used for image recognition and natural language processing (see, e.g., \cite{Soniya2015ARO}), but are being applied also to the financial market, e.g., to predict stock prices using textual news analysis (\cite{Day2016DeepLF}). Actually, the experiments reported in this paper confirm that DNNs obtain the best overall accuracy in the price forecast task under consideration, even if they require more time to be tuned, if compared to two state-of-the-art statistical models such as ARIMA and Prophet.

In this paper, we propose a framework that uses stock prices historical data to train a set of DNNs to forecast the future (next month) stock prices. Such predictions are exploited in a novel way to determine the most profitable technical indicator(s) to be used as the basis of the trading strategy that is then executed by a robot advisor. Indeed, typically, traders test their (algorithmic trading) strategies, i.e., the technical indicators to watch and how to react to their values, on the historical market data (the so called \textit{backtesting}), and then apply to the future trades the strategy that achieved the maximum profit on such past data. On the other hand, the term \textit{forward testing}  (also known as paper testing \footnote{\url{https://www.investopedia.com/terms/p/papertrade.asp}}) is sometimes used in the literature to indicate a technique that  updates the trading strategy in real time by looking at the current market data.
In this paper we propose a ''\textit{forward testing in the future}'', that we shall call \textit{DNN-forwardtesting} in the following, instead of the  real time one. In such a technique, the best strategy is devised by looking at the profits earned by applying the candidate strategies directly to the possible future predicted by the DNN. Indeed, we leverage on the capabilities of the DNN to learn from the past trends and characteristics that would be difficult or even impossible to capture using the simple indicator-based analyses performed by the classical approach. In such sense, our forward testing is able to better exploit the available historical data and allow a finer strategy definition. 

To verify our approach, we performed a model-based clustering using k-means++ on ten different stocks. Then we selected two shares, issued by companies operating in completely different sectors, from the cluster with the same medium volatility i.e., with price fluctuations that are not excessively large or small (as the ones, e.g., of a tech company stock or a large blue-chip company stock, respectively)

The experiments show that our technique allows the trader to choose a strategy that is more or equally profitable than the one that would be selected with traditional backtesting, if applied on the same historical data. Therefore, DNN-forwardtesting appears a better strategy selection criterion.
\section{Preliminaries}
\label{sec:prel}
This section describes some preliminary concepts related to trading, which are useful for a better understanding of the statistical processes adopted and the methodologies applied for trend price forecasting. 

Specifically, the first part will list the formulations of concepts related to technical analysis, cumulative returns and price volatility. In the second part, the performance evaluation metrics of the models and the trading strategies applied.

\subsection{Technical Analysis}
\label{sec:techAnalysis}
\begin{figure}[!ht]
	\centerline{\includegraphics[width=45em]{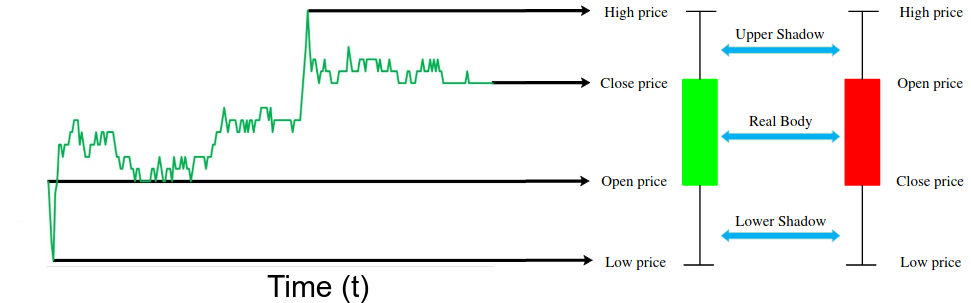}}
	\caption{Example of candlestick chart.}
	\label{fig:candlestick}
\end{figure}

Technical analysis (TA) represents the type of investment analysis that uses simple mathematical formulations or graphical representations of the time series of financial assets to explore trading opportunities. In its algorithmic form, TA uses the analysis of asset price history series (\cite{wang2021forecasting}), defined as OHLC, i.e., the opening, highers, lowest and closing prices of an asset, typically represented with candlesticks charts (see, e.g. in fig. \ref{fig:candlestick}). 

For each timeframe $t$, the OHLC of an asset is represented as a 4-dimensional vector
$X_t = (x^{(o)}_t,x^{(h)}_t,x^{(l)}_t,x^{(c)}_t)^T$, where $x_t^{(l)} > 0$, $x_t^{(l)} < x_t^{(h)}$ and  $x_t^{(o)}, x_t^{(c)} \in [x_t^{(l)},x_t^{(h)}]$.

\subsection{Cumulative Returns}
\label{sect:cumret}
Given a time series $TS$, the return of the $i$-th trade is defined as: $r_i = \frac{TS(sell_i)-TS(buy_i)}{TS(buy_i)}$ where $TS(sell_i)$ and $TS(buy_i)$ denote the price of conducting the $i$-th \textit{sell} and \textit{buy} trades, respectively (see \cite{YangCan2020Zhai}). Considering the transaction costs like \textit{fees \& slippage} (fs), a Cumulative Return (CR(n)) is the aggregate amount gained or lost over time with a set of trades, and it is calculated as follows:

$$CR(n)=\prod_{i=1}^n(1+\frac{TS(sell_i)-TS(buy_i)-fs_i}{TS(buy_i)})$$

\noindent where $n$ is the number of trades, and $fs_i$ the transaction costs in the $i$-th trade, which can be approximately considered as $fs_i \approx TS(s_i) \cdot k$, where $k$ is the transaction cost rate.

It is worth noting that the return in each sub-period of the $TS$ is rarely identical, so compounding it is impossible (as
detailed in \footnote{https://www.ivanletteri.it/2022/06/29/why-lr-is-better-then-sr/}). 

On the other hand, \textit{logaritmic cumulative returns} measure the rate of exponential growth, instead of measuring the percent of price change for each sub-period. It is measured the exponent of its natural growth during that time, and then it is added each $i$ sub-period of the exponential growth to get the total growth for the period, as follows:

$$ln(CR(n))=\sum_{i=1}^nln(1+\frac{TS(sell_i)-TS(buy_i)-fs_i}{TS(buy_i)}).$$

Logaritmic cumulative returns present the following three properties that can not be observed in raw prices (\cite{Thomas2000Mikosch}):
\begin{enumerate}
    \item The distribution of logaritmic cumulative returns is not normal, but rather with a high peak at zero and heavy-tailed. This comes from most days having little changes, i.e. a ``flat'' days.
    \item There is dependence in the tails of the distribution. These tails represent major changes in returns, which do not happen often, but tend to trend together\footnote{\url{https://digitalcommons.georgiasouthern.edu/etd/1409/}}. 
    \item The sample auto-covariance function (ACF) in the logaritmic return process is negligible at all lags, except possibly the first lag (\cite{Hoerlein2017ModelingVO}).
\end{enumerate}

For these reasons, in this work we focused on the logaritmic cumulative returns.

\subsection{Volatility}
\label{sect:volatility}

While forecasting changes in stock returns is a very hard task, forecasting the size of changes (i.e. the volatility) seems more promising. Volatility quantifies the dispersion of returns. Unfortunately, this dispersion can not be measured and volatility is not directly observable, but it is possible to estimate it (\cite{Pet2018Gab}).

Most finance textbooks use standard deviation (\textit{StdDev}) as a measure of volatility of a stock price as follows:

$$\sigma^2=\frac{1}{N}\sum_{1=1}^N(ln_{CR(n)}(x_t^{(c)})-ln_{CR(n)}(\bar{x}_t^{(c)}))^2$$

\noindent where  $ln_{CR(n)}(x_t^{(c)})$ are the logaritmic cumulative returns of the close price items, the $ln_{CR(n)}(\bar{x_t}^{(c)})$ are the mean values of these observations, and $N$ is the sample size. 

However, the use of standard deviation loses information on the ordering of the price movements, grouping them around the average value of the returns (\cite{Sen2021}).

For this reason (and others detailed in \footnote{https://www.ivanletteri.it/2022/06/29/why-stddev-is-not-a-good-measure-of-volatility/}), we focused on other formulations of historical volatility (HV) measures. 

The following types of \textbf{Historical Volatility} measures are specific and efficient in the Quantitative Finance context (``\textit{Measuring historical volatility}'' - Colin Bennett\footnote{\url{https://dynamiproject.files.wordpress.com/2016/01/measuring_historic_volatility.pdf}}).

\begin{itemize}
    \item The \textit{Parkinson} (PK) estimator incorporates the stock's daily \textit{high} and \textit{low} prices and is calculated as follow: 
    $$PK = \sqrt{\frac{1}{4Nln_{CR(n)}2}\sum_{i=1}^N(ln_{CR(n)}\frac{x_t^{(h)}}{x_t^{(l)}})^2}.$$
    
    PK provides completely separate information from using time-based sampling such as closing prices, but it cannot handle trends and jumps, so it systematically underestimates the volatility. 

    \item The \textit{Garman-Klass} (GK) estimator incorporates open, low, high, and close prices and is calculated as follows: 
    
    $$GK=\sqrt{\frac{1}{N}\sum_{i=1}^N \frac{1}{2}(ln_{CR(n)}\frac{x_t^{(h)}}{x_t^{(l)}})^2-\frac{1}{N}\sum_{i=1}^N(2ln_{CR(n)}2-1)(ln\frac{x_t^{(c)}}{x_t^{(o)}})^2}.$$
    This method is robust for opening jumps in price and trend movements. However, it takes into account not only the price at the beginning and end of the time interval but also intraday price extremums.
    
    \item The \textit{Rogers-Satchell} (RS) estimator, unlike the PK and GK estimators, RS incorporates drift term (mean return not equal to zero). It is calculated as follows:
    
    $$RS = \sqrt{\frac{1}{N}\sum_{t=1}^{N}(ln_{CR(n)}(\frac{x_t^{(h)}}{x_t^{(c)}})ln_{CR(n)}(\frac{x_t^{(h)}}{x_t^{(o)}})+ln_{CR(n)}(\frac{x_t^{(l)}}{x_t^{(c)}})ln_{CR(n)}(\frac{x_t^{(l)}}{x_t^{(o)}})}.$$
    
    However, it does not take into account price movements between trading sessions, and underestimate volatility since price jumps periodically occur between sessions.
    
    \item The \textit{Yang-Zhang} (YZ) estimator \cite{Yang2000Zhang} incorporates open, low, high, and close prices, and is calculated as follows: 
    
    $$\sigma_{OpentoCloseVol}^2 = \frac{1}{N-1}\sum_{i=1}^N(ln_{CR(n)}\frac{x_t^{(c)}}{x_t^{(o)}} - ln_{CR(n)}\frac{x_t^{(c)}}{x_t^{(o)}})^2$$
    $$\sigma_{OvernightVol}^2 = \frac{1}{N-1}\sum_{i=1}^N(ln_{CR(n)}\frac{x_t^{(o)}}{x_{t-1}^{(c)}} - ln_{CR(n)}\frac{x_t^{(o)}}{x_{t-1}^{(c)}})^2$$
    $$YZ = \sqrt{\sigma_{OvernightVol}^2 +k\sigma_{OpentoCloseVol}^2+(1-k)\sigma^2_{RS}}$$
    
    \noindent where $k = \frac{0.34}{1.34+\frac{N+1}{N-1}}$.
    
    YZ is considered the most powerful volatility estimator. It has the minimum estimation error since it is a weighted average of RS, the close-open volatility and the open-close volatility.
\end{itemize}

\subsection{Trading Strategies}

In this section, we discuss two distinct trading strategy classes used in our framework: \textit{trend following} and \textit{mean reversion}. 

\begin{figure}[ht]
	\centerline{\includegraphics[width=30em]{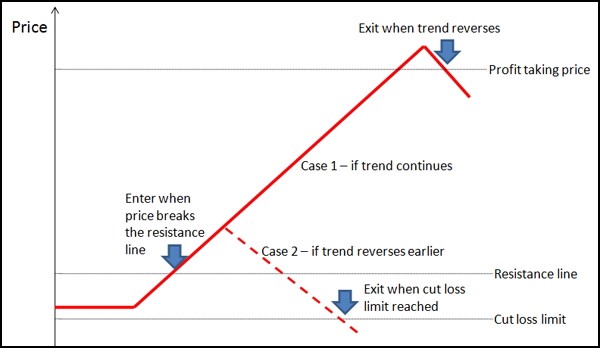}}
	\caption{Trend Following Strategy taken from \url{trendfollowing.com}}
	\label{fig:trendFollow}
\end{figure}

\textbf{Trend Following.} One way to trade a trend is to look at an asset with a resistance line (see fig. \ref{fig:trendFollow}). Once the price breaks through resistance, a trader places an order in the direction of the breakout\footnote{https://en.wikipedia.org/wiki/Breakout\_(technical\_analysis)}.

Trend-following, or momentum, strategies have the attractive property of generating trading returns with a positively skewed statistical distribution. Consequently, they tend to hold on to their profits and are unlikely to have severe `drawdowns' (\cite{Martin2021Richard}). 

\begin{figure}[ht]
	\centerline{\includegraphics[width=35em]{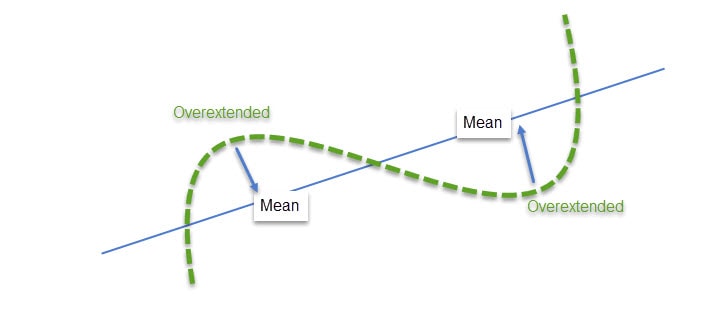}}
	\caption{Mean Reversion Strategy description from \url{tradergav.com/what-is-mean-reversion-trading-strategy/}}
	\label{fig:meanReversion}
\end{figure}

\textbf{Mean Reversion.} The idea of mean reversion strategies is that the maximum and minimum price of a security is temporary, and that it will tend to the average over time (see \cite{huang2016profitability}). 

Fig. \ref{fig:meanReversion} shows how the mean reversion is applied by first identifying the stock's trading range and then calculating the average price using analytical techniques including volatility (\ref{sect:volat_analisys}). 

\cite{Robert2005Elliott} explains how mean-reverting processes might be used in pairs trading and developed several methods for parameter estimation. \cite{Marco2010Avellaneda} used mean-reverting processes for pairs trading, and modeled the hitting time to find the exit rule of the trade.

It is worth noting that trend following and mean reversion strategies, although theoretically opposing ideas, are not in conflict with each other and they are therefore both applicable at the same time to the same security. Obviously, it is not possible to apply them to the same time window in order to make money with both, but there is no reason why they are incompatible with each other. Trends, in fact, tend to occur on longer time horizons, while reversals, which are in fact small swings around these trends, on shorter ones. Therefore, in our tests we used them both alternatively.

\subsection{Forecasting Error Metrics}
In this section we introduce the different metrics used to estimate the forecasting error of the models presented in the paper.

\begin{itemize}
    \item \textit{Mean Square Error} (MSE): measures the average of the squared difference between the correct and predicted values (called prediction \textit{error} or \textit{residual}) as follows:
    
	$$MSE(y,\hat{y})=\sum_{t=0}^N(y_t-\hat{y_t})^2$$
 
    \item \textit{Root Mean Square Error} (RMSE): is the standard deviation of the residuals (prediction errors), i.e, 
    how spread out these residuals are, as follows: 
    
    $$RMSE(y,\hat{y})=\sqrt{\sum_{t=0}^N(y_t-\hat{y_t})^2}.$$
    
    \item \textit{Mean Absolute Error} (MAE): measures the average of the absolute differences between the correct and predicted values as follows:
    
    $$MAE(y,\hat{y})=\frac{1}{N}\sum_{t=0}^N|\frac{y_t-\hat{y_t}}{y_t}|$$
    
    \noindent where $n$ is the number of fitted points, $y$ is the actual value, $\hat{y}$ is the forecast value.
	
    \item \textit{Mean Absolute Percentage Error} (MAPE): measures the accuracy as a percentage, and can be calculated as the average absolute percent error for each time period minus the actual values divided by the actual values. 
    MAPE is the most common measure used to forecast error, and works best if there are no extremes to the data (and no zeros).

    \item \textit{Explained Variance Regression Score} (EVS): measures the error dispersion as follows:
    
	$$EVS(y,\hat{y})=1-\frac{Var(\hat{y}-y)}{Var(y)}$$
	
    \noindent where the $Var$ is biased variance, (i.e. $Var(\hat{y}-y)=\frac{1}{n}\sum(error-mean(error))$.
\end{itemize}

\subsection{Profit and Risk Metrics} 
In this section, we set out the metrics related to downside risk by estimating the potential loss in value of the stock.

\begin{itemize}
    \item The \textit{Maximum drawdown (MDD)}\footnote{https://www.investopedia.com/terms/m/maximum-drawdown-mdd.asp} measures the largest decline from the peak in the whole trading period, to show the worst case, as follows: $MDD=max_{\tau \in (0,t)}[max_{t \in (0,\tau)}\frac{n_t-n_{\tau}}{n_t}]$.
    \item The \textit{Sharpe ratio (SR)}\footnote{https://www.investopedia.com/terms/s/sharperatio.asp} is a risk-adjusted profit measure, which refers to the return per unit of deviation as follows: $SR = \frac{\mathbb{E}[r]}{[r]}$.
    \item The \textit{Sortino ratio (SoR)}\footnote{https://www.investopedia.com/terms/s/sortinoratio.asp} is a variant of the risk-adjusted profit measure, which applies DD as risk measure: $SoR = \frac{\mathbb{E}[r]}{DD}$.
    \item The \textit{Calmar ratio (CR)}\footnote{https://www.investopedia.com/terms/c/calmarratio.asp} is another variant of the risk-adjusted profit measure, which applies MDD as risk measure: $CR = \frac{\mathbb{E}[r]}{MDD}$.
\end{itemize}

To check the goodness of trades, we mainly focused on the \textit{Total Returns} $R_{k}(t)$ for each stock $(k = 1, ...,p)$ in the time interval $(t= 1, ...,n)$ 

\begin{equation}
    R_{k}(t) = \frac{Z_k(t+\Delta t) - Z_{k}(t)}{Z_{k}(t)}
\end{equation}

and furthermore analysing the standardized returns $r_k = (R_k - \mu_k) / \sigma_k,$ with $(k = 1, ...,p)$, where $\sigma_k$ is the standard deviation of $R_k$, e $\mu_k$ denote the average overtime for the studied period.

\section{Dataset Analysis}
\label{sec:dataset}
\label{sec:methodology}

This section describes the methodology through which the final dataset has been collected for the experiment, the criteria adopted for the statistical analysis of the data, the training process and the tests conducted on the DNNs with optimized processes (\cite{LetteriDSopt2020}). 

We started by analysing ten stocks picked randomly among the securities listed on the New York Stock Exchange (NYSE) shown in tab. \ref{tab:vol_feats}.

\begin{table*}[!th]
\centering
\caption{List of 10 stocks randomly selected  [source \url{it.finance.yahoo.com}].}
\label{tab:vol_feats}
\begin{tabular}{|l|l|l|}
\hline
\multicolumn{1}{|c|}{\textbf{Ticker}} & \multicolumn{1}{c|}{\textbf{Company}} & \multicolumn{1}{c|}{\textbf{Market}} \\ \hline
CSGKF                                 & \multicolumn{1}{c|}{Credit Suisse Group AG}        & \multicolumn{1}{c|}{Other OTC}             \\ \hline
EOG                                   & \multicolumn{1}{c|}{EOG Resources, Inc.}       & \multicolumn{1}{c|}{NYSE}             \\ \hline
META                                  & \multicolumn{1}{c|}{Meta Platforms, Inc. }        & \multicolumn{1}{c|}{Nasdaq GS}             \\ \hline
NKE                                  & \multicolumn{1}{c|}{NIKE, Inc.}        & \multicolumn{1}{c|}{NYSE}             \\ \hline
DIS                                   & \multicolumn{1}{c|}{The Walt Disney Company}                       &   \multicolumn{1}{c|}{NYSE}                                    \\ \hline
PG                                    &  \multicolumn{1}{c|}{The Procter \& Gamble Company}                                     & \multicolumn{1}{c|}{NYSE}                                      \\ \hline
QQQ                                   &  \multicolumn{1}{c|}{Invesco QQQ Trust}                             & \multicolumn{1}{c|}{Nasdaq GM}                                     \\ \hline
IBM                                   &  \multicolumn{1}{c|}{International Business Machines Corporation}                                     & \multicolumn{1}{c|}{NYSE}                                     \\ \hline
ANF                                   &   \multicolumn{1}{c|}{Abercrombie \& Fitch Co.}                                    & \multicolumn{1}{c|}{NYSE}             \\ \hline
CS                                    &  \multicolumn{1}{c|}{Credit Suisse Group AG}                                    &    \multicolumn{1}{c|}{NYSE}                                  \\ \hline
\end{tabular}
\end{table*}

Technically speaking, the dataset used in this paper consists of the time series of OHLC prices, over the time period from 30 October, 2011 to 30 November, 2021, for a total period of 2537 open market days. OHLC prices are the opening, highest, lowest and closing prices of an asset, and are commonly used to analyze the assets price history when performing the so called technical analysis (see sect. \ref{sec:techAnalysis}) to explore trading opportunities.

\subsection{Volatility Analysis}
\label{sect:volat_analisys}

Some stocks are more volatile than others. For example, the shares of a large blue-chip company may not have large price fluctuations and are therefore said to have \textit{low volatility}, whereas the shares of a tech stock which fluctuate often having \textit{high volatility}. There are also stocks with \textit{medium volatility}\footnote{https://www.fool.com/investing/how-to-invest/stocks/stock-market-volatility/}, and this is the case of the assets that we decided to select in this paper.

Therefore, in order to filter stocks with medium volatility, we first created a dataset composed by the time series of the PK, GK, RS, YZ historical volatility estimators for out securities. The data has been standardazed and clustered with K-means++ \cite{Arthur2007Vassilvitskii}.

The procedure adopted to find the optimal number of clusters is the Elbow method  \footnote{\url{https://www.ivanletteri.it/2022/10/11/optimal-volatility-clusters/}}, i.e.:

\begin{enumerate}
    \item Compute k-means clustering for different values of $k$. In our case, we varied $k$ from 2 to 20 clusters.
    \item For each $k$, is calculated the total within-cluster sum of squares (wss).
    \item Plot the curve of wss according to the number of clusters $k$.
    \item Find the location of a bend (knee) in the plot, which is generally considered as indicator of the appropriate number of clusters.
\end{enumerate}

\begin{figure}[!ht]
	\centerline{\includegraphics[width=40em]{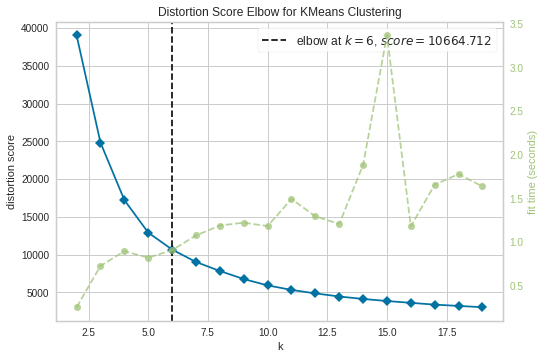}}
	\caption{Elbow method to determine optimal cluster in $k=6$}
	\label{fig:elbow_method}
\end{figure}

\begin{figure}[!ht]
	\centerline{\includegraphics[width=45em]{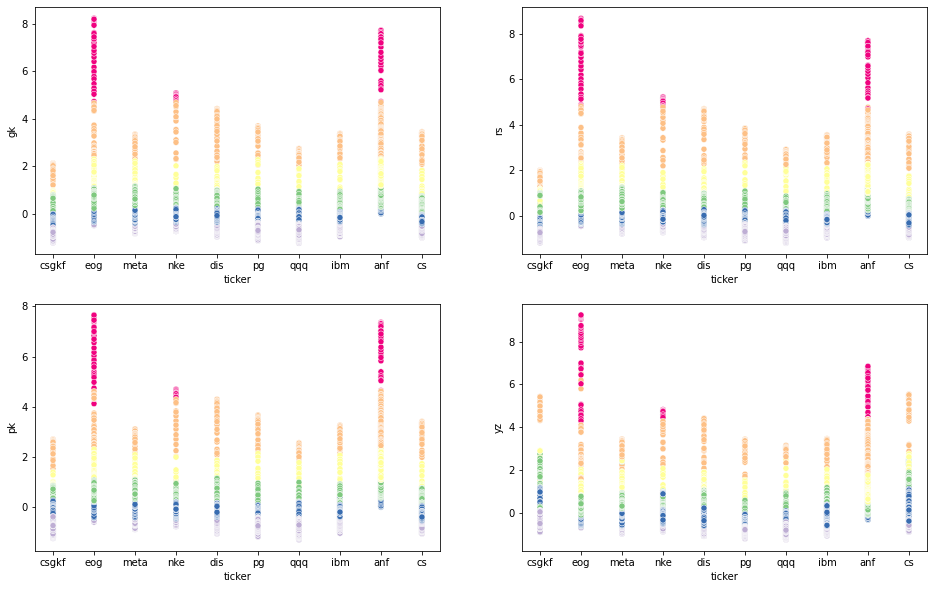}}
	\caption{Kmeans++ Clusters with $k=6$ of the Historical Volatility estimators dataset. }
	\label{fig:kmeanClusters}
\end{figure}

Fig. \ref{fig:elbow_method} shows that the best clustering is for $k=6$, and fig. \ref{fig:kmeanClusters} clearly shows that, between the considered stocks, ANF and EOG have all the types of observations spread over all the k-means clusters foreach HV estimators considered. This makes them good candidates for our final dataset because they gather all the characteristics (pros and cons) of the four history volatility estimators adopted. 

\begin{table}[!ht]
\centering
\caption{Volatility measures of the HV estimators.}
\label{tab:volMetrEstims}
\begin{tabular}{ccllllcllll}
\cline{2-5} \cline{8-11}
\multicolumn{1}{c|}{\textbf{}}                                                                & \multicolumn{4}{c|}{\textbf{ANF Volatility Measures}}                                                                                             &                       & \multicolumn{1}{l|}{}                                                                        & \multicolumn{4}{c|}{\textbf{EOG Volatility Measures}}                                                                                             \\ \cline{1-5} \cline{7-11} 
\multicolumn{1}{|l|}{\textbf{\begin{tabular}[c]{@{}l@{}}HV\\ Est.\end{tabular}}} & \multicolumn{1}{c|}{\textit{Min}} & \multicolumn{1}{c|}{\textit{Max}} & \multicolumn{1}{c|}{\textit{Mean}} & \multicolumn{1}{l|}{\textit{StdDev}} & \multicolumn{1}{l|}{} & \multicolumn{1}{l|}{\textbf{\begin{tabular}[c]{@{}l@{}}HV\\ Est.\end{tabular}}} & \multicolumn{1}{c|}{\textit{Min}} & \multicolumn{1}{c|}{\textit{Max}} & \multicolumn{1}{c|}{\textit{Mean}} & \multicolumn{1}{c|}{\textit{StdDev}} \\ \cline{1-5} \cline{7-11} 
\multicolumn{1}{|c|}{PR}                                                                      & \multicolumn{1}{c|}{-0.264694}    & \multicolumn{1}{c|}{0.344451}     & \multicolumn{1}{l|}{0.000448}      & \multicolumn{1}{l|}{0.034559}        & \multicolumn{1}{l|}{} & \multicolumn{1}{c|}{PR}                                                                      & \multicolumn{1}{l|}{-0.320072}    & \multicolumn{1}{l|}{0.165702}     & \multicolumn{1}{l|}{0.000625}      & \multicolumn{1}{l|}{0.024650}        \\ \cline{1-5} \cline{7-11} 
\multicolumn{1}{l}{}                                                                          & \multicolumn{1}{l}{}              &                                   &                                    &                                      &                       & \multicolumn{1}{l}{}                                                                         &                                   &                                   &                                    &                                      \\ \cline{1-5} \cline{7-11} 
\multicolumn{1}{|c|}{\textit{PK}}                                                            & \multicolumn{1}{c|}{0.205809}     & \multicolumn{1}{c|}{1.061728}     & \multicolumn{1}{c|}{0.401217}      & \multicolumn{1}{l|}{0.116944}        & \multicolumn{1}{l|}{} & \multicolumn{1}{c|}{\textit{PK}}                                                            & \multicolumn{1}{l|}{0.136258}     & \multicolumn{1}{l|}{1.096084}     & \multicolumn{1}{l|}{0.290084}      & \multicolumn{1}{l|}{0.122853}        \\ \cline{1-5} \cline{7-11} 
\multicolumn{1}{|c|}{\textit{GK}}                                                           & \multicolumn{1}{c|}{0.204787}     & \multicolumn{1}{l|}{1.128780}     & \multicolumn{1}{c|}{0.406463}      & \multicolumn{1}{l|}{0.121297}        & \multicolumn{1}{l|}{} & \multicolumn{1}{c|}{\textit{GK}}                                                           & \multicolumn{1}{l|}{0.145601}     & \multicolumn{1}{l|}{1.188479}     & \multicolumn{1}{l|}{0.291579}      & \multicolumn{1}{l|}{0.128462}        \\ \cline{1-5} \cline{7-11} 
\multicolumn{1}{|c|}{\textit{RS}}                                                           & \multicolumn{1}{c|}{0.203886}     & \multicolumn{1}{l|}{1.146890}     & \multicolumn{1}{l|}{0.409334}      & \multicolumn{1}{l|}{0.123215}        & \multicolumn{1}{l|}{} & \multicolumn{1}{c|}{\textit{RS}}                                                           & \multicolumn{1}{l|}{0.148818}     & \multicolumn{1}{l|}{1.266812}     & \multicolumn{1}{l|}{0.292796}      & \multicolumn{1}{l|}{0.134589}        \\ \cline{1-5} \cline{7-11} 
\multicolumn{1}{|c|}{\textit{YZ}}                                                           & \multicolumn{1}{c|}{0.236222}     & \multicolumn{1}{l|}{1.521485}     & \multicolumn{1}{l|}{0.527025}      & \multicolumn{1}{l|}{0.194645}        & \multicolumn{1}{l|}{} & \multicolumn{1}{c|}{\textit{YZ}}                                                           & \multicolumn{1}{l|}{0.169166}     & \multicolumn{1}{l|}{1.952395}     & \multicolumn{1}{l|}{0.366873}      & \multicolumn{1}{l|}{0.210179}        \\ \cline{1-5} \cline{7-11} 
\end{tabular}
\end{table}

\cite{BlackBESS1976} observed that the volatility tends to increase when the price drops and appears a significant skew in the volatility. As shown in tab. \ref{tab:volMetrEstims}, we noted that the percentage returns (PR) respect to the volatility measures, provided by all the estimators, consistently suggests that it may be more useful in estimating the market volatility for short-term trading purposes rather than characterizing the evolution of the historical
volatility over the long term.

\newpage

\subsection{ANF and EOG price time series}
ANF was founded at the end of September 1996, and from April to October 2011 it was several times close to the all-time high, always encountering resistance. 

\begin{figure}[!ht]
\centerline
 {\includegraphics[width=0.99\hsize]{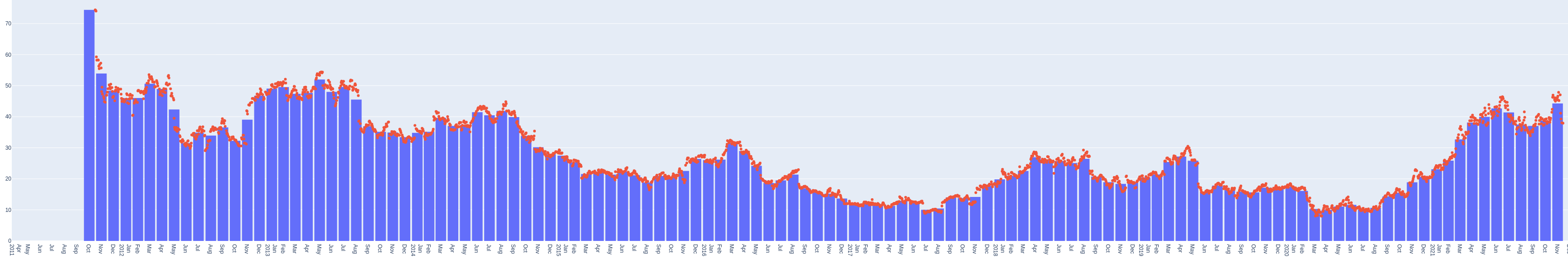}}
	\caption{ANF average trend from October 30th, 2011 to November 30th, 2021}
	\label{fig:ANF_trend}
\end{figure}

On November 23, 2021 the company CEO announced net sales of \$905 million, up 10\% as compared to the previous year and up 5\% as compared to the pre-COVID 2019 third quarter net sales (source GlobeNewswire\footnote{www.globenewswire.com/news-release/2021/11/23/2339734/0/en/Abercrombie-Fitch-Co-Reports-Third-Quarter-Results.html}).  Fig. \ref{fig:ANF_trend}  shows the ANF stock price trends starting from October 30th, 2011 to November 30th, 2021.
On the other hand, the EOG stock, with a market value of \$55.21 billion and 84.08\% institutional ownership, has gained 9.39\% so far (source investopedia\footnote{https://www.investopedia.com/3-oil-and-gas-stocks-to-watch-this-week-4628357}). 
\begin{figure}[ht]
	\centerline{\includegraphics[width=0.99\hsize]{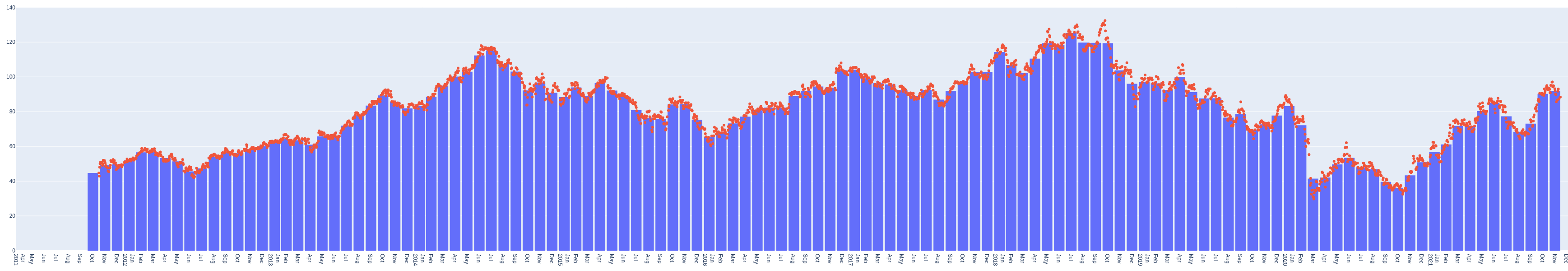}}
	\caption{EOG average trend from October 30th, 2011 to November 30th, 2021}
	\label{fig:EOG_trend}
\end{figure}
The company is expected to post quarterly earnings of \$3.24 per share in its next report. Fig. \ref{fig:EOG_trend}  shows the EOG stock price trends in the same time interval used for ANF.

ANF and EOG are certainly assets with a sometimes controversial trend and consequently well profitable if rightly analyzed, especially in the particular period of 2020/2021 due to the global pandemic. However, in the analysed period, ANF and EOG do not show as the classic always profitable stocks (e.g. Tesla, Apple, Microsoft, or Bitcoin) to which trivially apply, in that period, a passive Buy and Hold strategy\footnote{https://www.investopedia.com/terms/b/buyandhold.asp} for gains.

The time series of price observations can be downloaded from Yahoo Finance\footnote{https://it.finance.yahoo.com/quote/ANF?p=ANF\&.tsrc} and \footnote{https://it.finance.yahoo.com/quote/EOG?p=EOG\&.tsrc}, or via our github repository\footnote{\url{https://github.com/IvanLetteri/DNN-ForwardTesting}} also contains a copy of such data preprocessed and split into train and test sets to be used in a deep neural network.

\subsection{Synchrony between ANF and EOG stocks}
In order to prove that the dataset is general enough to model a variety of different shares with more or less the same volatility coefficient ($vc$), we also have to prove that the price time series corresponding to the selected assets are completely uncorrelated, i.e., ANF and EOG does not influence each other.

Therefore, after scaling the values with a  \textit{minMax} normalization ($x_i=\frac{x-x_{min}}{x_{max}-x_{min}}$), we evaluated the \textit{synchrony} between the two financial assets using the Pearson coefficient and the Dynamic Time Warping.

The Pearson coefficient measures the linear relation between two continuous signals, and is defined as $r = \frac{\sum_i(x_i-\bar{x})(y_i-\bar{y})}{\sqrt{\sum_i(x_i-\bar{x})^2}\sqrt{\sum_i(y_i-\bar{y})^2}}$, where $x_i$ and $y_i$ are, in our case, the close prices of the ANF and EOG stocks.

\begin{figure}[ht]
	\centerline{\includegraphics[width=0.5\hsize]{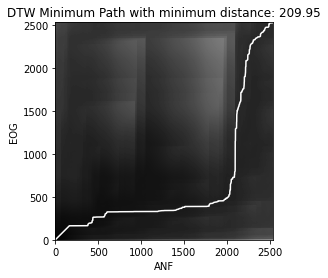}}
	\caption{DTW Minimum Path with minimum distance between ANF and EOG close prices.}
	\label{fig:dtw}
\end{figure}

In fig. \ref{fig:dtw}(b), the coefficients -1 and 1 indicate a perfect negative and positive correlation, respectively, whereas 0 stands for no correlation. It is worth noting that the Pearson coefficient is highly sensible to outliers, which can sensibly alter the correlation estimation, and requires the compared time series to contain homoscedastic data, having an homogeneous variance in the observation interval. However, both the considered time series do not contain many outliers and can be considered homoscedastic thanks to the medium volatility of the considered stock prices.

The overall Pearson coefficient between ANF and EOG is $0.28$, 
which confirms that the two stocks are almost completely uncorrelated. However, this is a measure of the \textit{global synchrony} in the overall period.
\begin{figure}[ht]
	\centerline{\includegraphics[width=1.0\hsize]{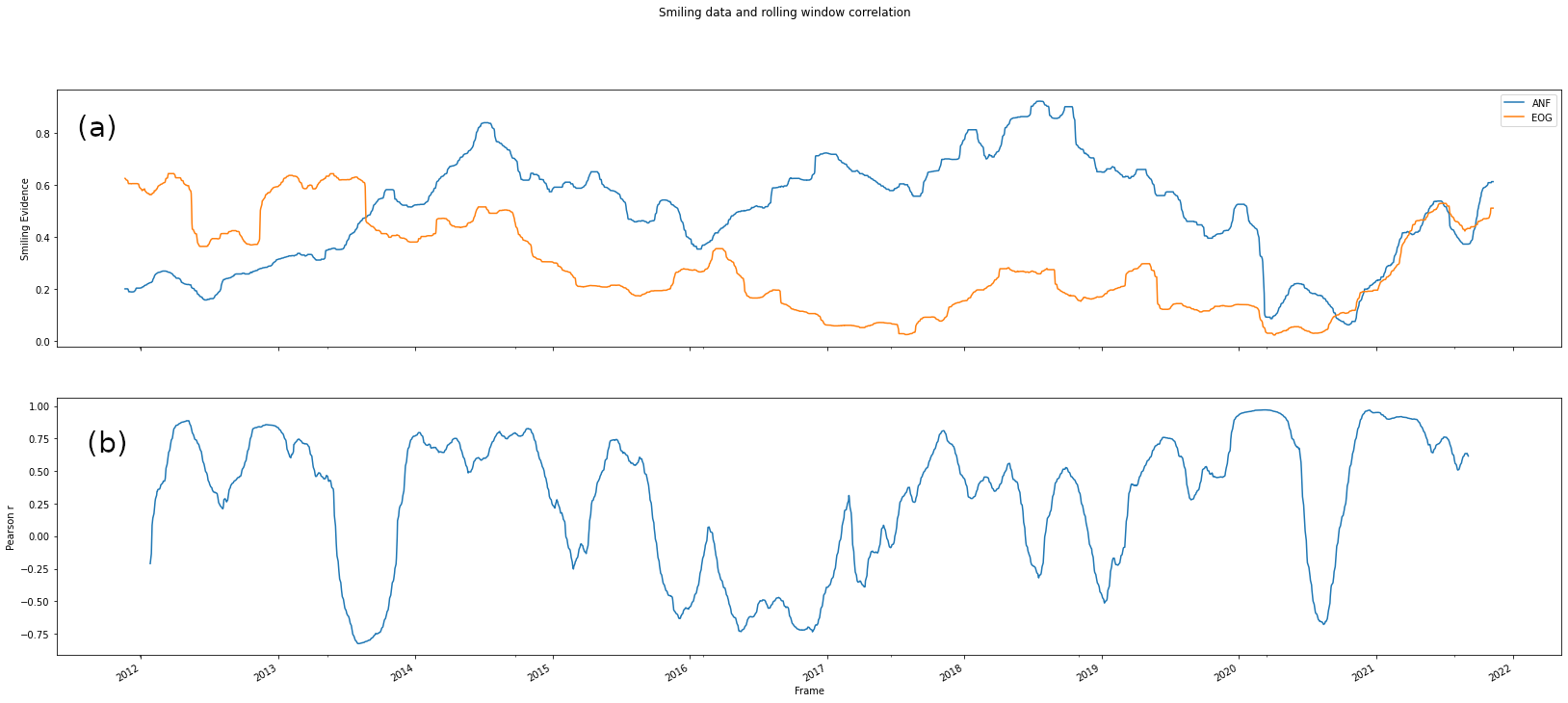}}
	\caption{Figure (a) show similing data between ANF and EOG, figure (b) the correlation coefficients.}
	\label{fig:EOG-ANF-PearsonCorr}
\end{figure}

Therefore, for sake of completeness, we calculated the \textit{local synchrony} in a small portions of the timespan by repeating the process along a moving window of $120$ samples. Fig. \ref{fig:EOG-ANF-PearsonCorr}(b) plots such moment-by-moment synchrony curve.

The Dynamic Time Warping (DTW) algorithm outperforms the Pearson correlation in detecting atypical functional dependencies ( \cite{LINKE2020117383}) between time series, even if they have a different number of samples. It calculates the optimal match between the two series by minimising the Euclidean distance between pairs of samples at the same time.

In Fig. \ref{fig:dtw}, we can see the optimal match between the closing price of the ANF and EOG assets. The minimum path cost is $d = 209.95$, and such a large distance between the two stocks supports our hypothesis of a complete absence of influence between them.

\newpage

\section{Forecasting with Statistical Methods}
\label{sec:statistical}
As a baseline for the proposed DNN-based forecast process, in this section we evaluate the performances of two well-known statistical time series forecasting methods, i.e., the ARIMA autoregressive model and the Prophet procedure, applied to our dataset.


\subsection{ARIMA model}
\label{sect:arima}
In this work, we applied a non-seasonal ARIMA models, neglecting the modulation effects of holidays and using therefore pure trend lines.

In general, an ARIMA model needs three parameters to run: the number of autoregressive terms $p$, the number of nonseasonal differences needed for stationarity $d$, and the number of lagged forecast errors in the prediction equation $q$ (see, e.g., \cite{Ham94} for more information on the meaning of such parameters).

As a first step, we checked the stationarity properties of the time series performing the Augmented Dikey Fuller (ADF) unit root test using the built-in method in the \textit{statsmodels} Python package. 

\begin{table}[!ht]
	\centering
	\caption{ADF test stationarity with AIC optimization.}
	\label{tab:adf-test}
	\begin{tabular}{l|c|c|c|c|ccc|}
		\cline{2-8}
		& \textbf{Test Statistic} & \textbf{p-value} & \textbf{Lags} & \textbf{Observations} & \multicolumn{3}{c|}{\textbf{Critical Value}}                                                  \\ \cline{6-8} 
		&                                          &                                   &                                &                                        & \multicolumn{1}{c|}{\textbf{1\%}} & \multicolumn{1}{c|}{\textbf{5\%}} & \textbf{10\%}         \\ \hline
		\multicolumn{1}{|l|}{\textit{ANF}} & -2.302                                   & 0.171                             & 5                              & 2529                                   & \multicolumn{1}{c|}{-3.432}       & \multicolumn{1}{c|}{-2.863}       & -2.567                \\ \hline
		\multicolumn{1}{|l|}{\textit{EOG}} & -2.422                                  & 0.135                             & 5                              & 2529                                   & \multicolumn{1}{c|}{-3.433}       & \multicolumn{1}{c|}{-2.862}       & -2.567                \\ \hline
		
	\end{tabular}
\end{table}

Tab. \ref{tab:adf-test} shows the number of lags considered, automatically selected based on the Akaike Information Criterion (AIC) (see \cite{Akaike1974H}) on ANF and EOG stock prices. The p-value results above the threshold (such as 5\% or 1\%) and suggests rejecting the null hypothesis, so the time series turns out not to be stationary, so the \textit{Volatility} is not dependent on time/trend.

The results achieved exposed in tab. \ref{tab:aic_arima} are qualitatively similar for all the two assets selected: for the $log(p_t)$ time series one cannot reject the null hypothesis of the presence of a unit root, signalling the non-stationarity of the series but with a first differencing the series, i.e. considering $\Delta log(p_t) = log(p_t) - log(p_{t-1})$, makes it stationary. Thus the $log(p_t)$ series are integrated of order one, and accordingly the models we considered are $ARIMA(p, 1, q)$.

In order to have a rough indication on the AR orders $p$, and on the MA orders $q$, we computed the sample autocorrelation function (ACF) and the partial autocorrelation function (PACF) for
$\Delta log(p_t)$ where:
\begin{itemize}
    \item for an exact $MA(q)$, the ACF is zero when lags larger than $q$; 
    \item for an exact $AR(p)$, the PACF is zero when lags larger than $p$.
\end{itemize}

\begin{figure}[ht]
	\centerline{\includegraphics[width=1.0\hsize]{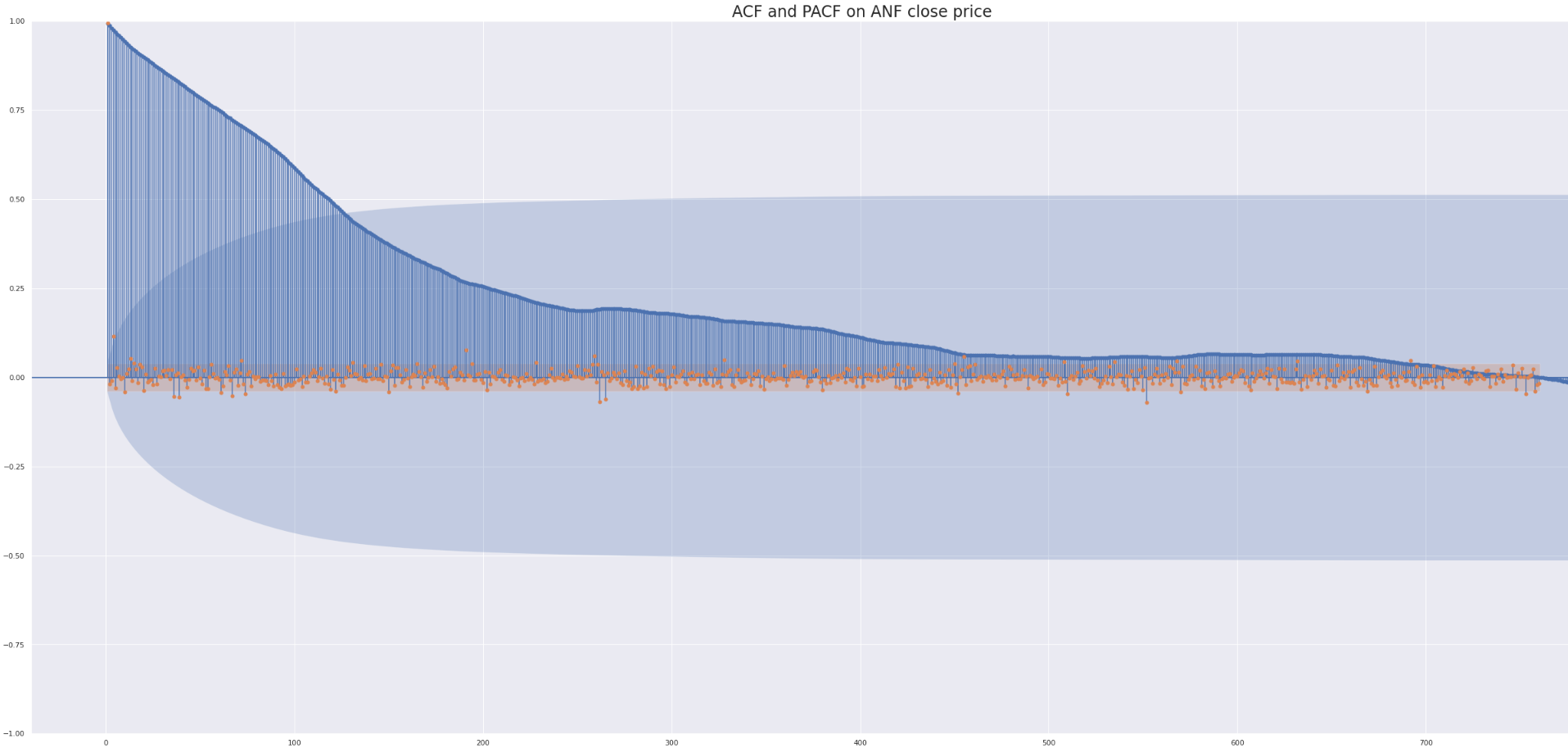}}
	\caption{ACF and PACF of the ANF stock closing price.}
	\label{fig:ANF_acf_pacf}
\end{figure}

\begin{figure}[ht]	\centerline{\includegraphics[width=0.99\hsize]{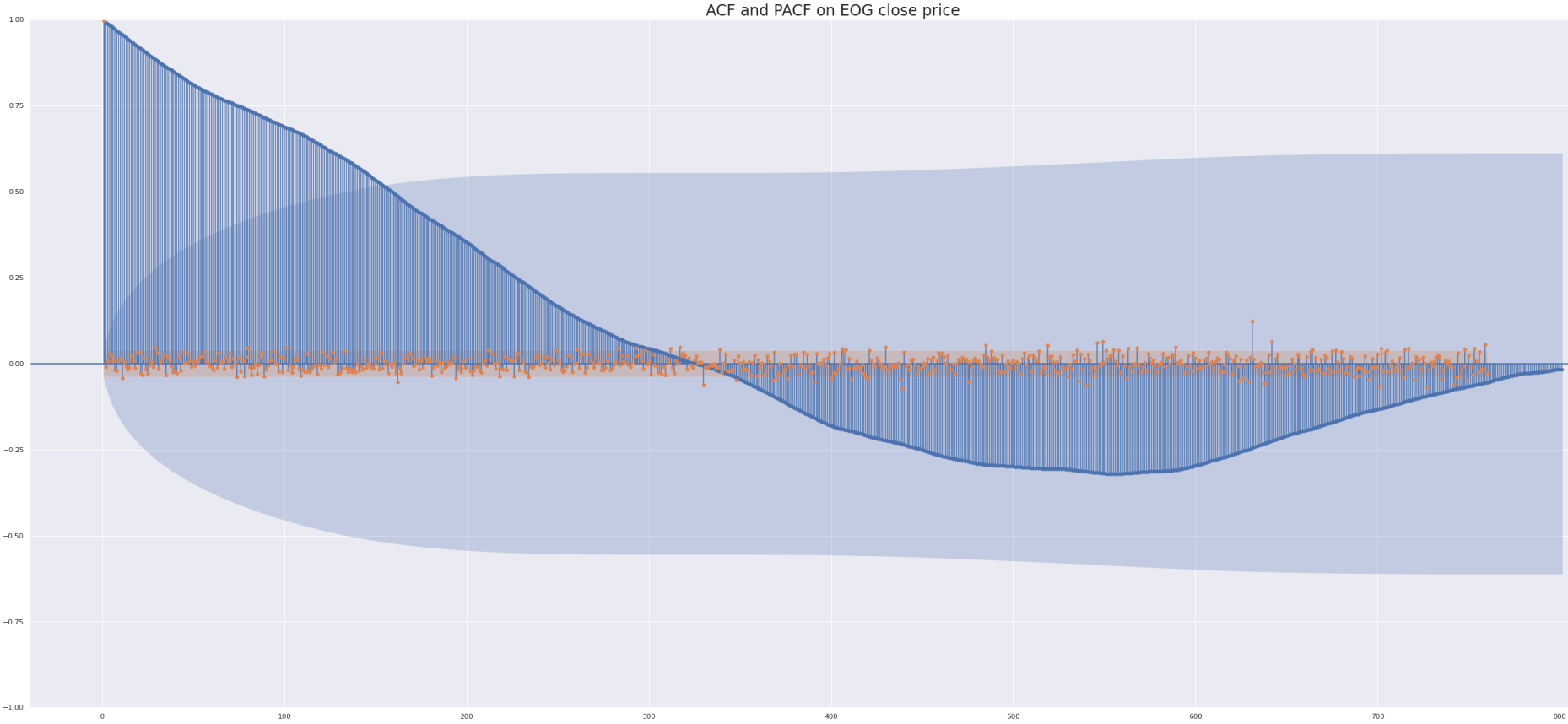}}
	\caption{ACF and PACF of the EOG stock closing price.}
	\label{fig:EOG_acf_pacf}
\end{figure}

\begin{figure}[ht]	\centerline{\includegraphics[width=0.99\hsize]{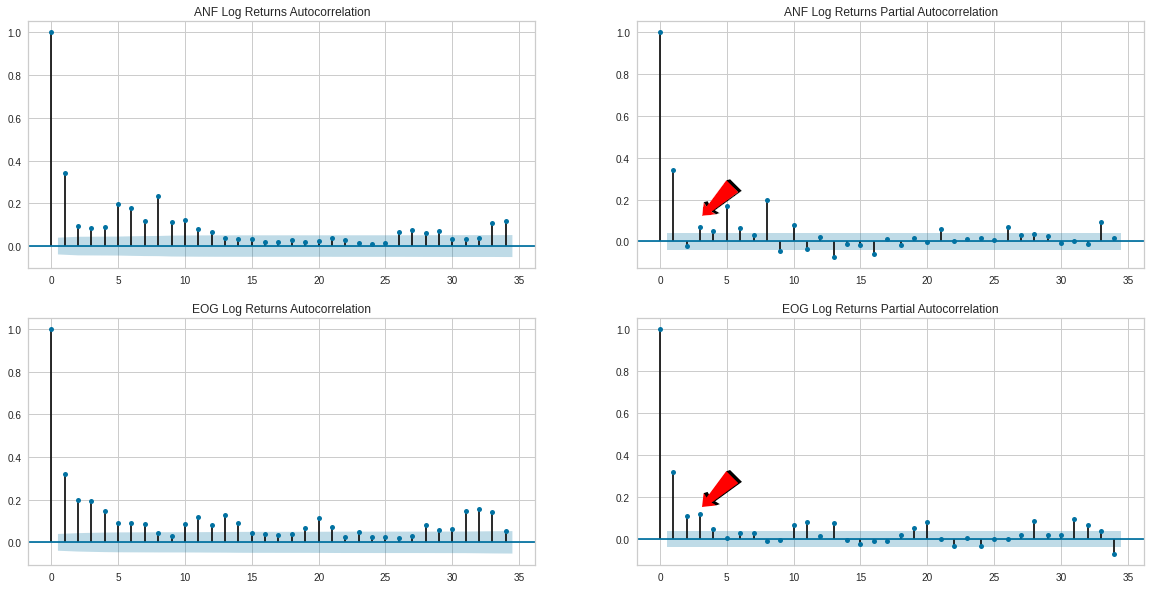}}
	\caption{ACF and PACF of the ANF and EOG stock closing prices.}
	\label{fig:ANF-EOG_acf_pacf}
\end{figure}

In our case, we can see from the ACF plot in Fig. \ref{fig:ANF_acf_pacf}, the plot of lag values on the x-axis and how they increase, and at the same time the correlation between price and lagged price on the y-axis deteriorates to lag = 750 for ANF and lag = 330 for EOG (see fig. \ref{fig:ANF-EOG_acf_pacf}). This means that the values of the historical series of ANF are strongly correlated with those of the lagged series for only an initial period, then the correlation decreases faster and faster, especially in case of EOG stock.

One of the most widely used algorithms is the auto.arima() function developed for automatic time series forecasting with ARIMA models (\cite{Hyndman2008Khandakar}). More precisely, we used the Auto-ARIMA model implemented in the \textit{pmdarima} \footnote{https://alkaline-ml.com/pmdarima/setup.html} library, it discovers the optimal parameters trying various sets of $p$ and $q$ to minimize the selected criterion that is, in our case, AIC, since it provides a good trade-off between the model fitting and the evaluation simplicity (\cite{Stoica2004Selen}) and also deals with the risk of overfitting and underfitting. The lower is the AIC value, the better is the result.

Despite that this algorithm allow us to implement the order selection process with relative ease in a standalone computer for short time series, efficient ARIMA model selection for ultra-long time series like our dataset is challenging, also with modern distributed computational environments like Google Colab. So, from the results of ACF and PACF shown in figs. \ref{fig:ANF-EOG_acf_pacf}, in order to optimize we established a range to analize for the $p$ value from $0$ to $5$ and for the $q$ value from $0$ to $2$.

\begin{table}[!ht]
\centering
\caption{ARIMA (p,d,q) models of ANF and EOG performing stepwise search to minimize AIC.}
\label{tab:aic_arima}
\begin{tabular}{cclcc}
\cline{1-2} \cline{4-5}
\multicolumn{1}{|c|}{\textbf{\begin{tabular}[c]{@{}c@{}}ARIMA\\ ( p,d,q )\end{tabular}}} & \multicolumn{1}{c|}{\textbf{\begin{tabular}[c]{@{}c@{}}ANF\\ AIC\end{tabular}}} & \multicolumn{1}{l|}{} & \multicolumn{1}{c|}{\textbf{\begin{tabular}[c]{@{}c@{}}ARIMA\\ ( p,d,q )\end{tabular}}} & \multicolumn{1}{c|}{\textbf{\begin{tabular}[c]{@{}c@{}}EOG\\ AIC\end{tabular}}} \\ \cline{1-2} \cline{4-5} 
\multicolumn{1}{|c|}{( 0,1,0 )}                                                          & \multicolumn{1}{c|}{8323.575}                                                   & \multicolumn{1}{l|}{} & \multicolumn{1}{c|}{\textbf{( 0,1,0 )}}                                                 & \multicolumn{1}{c|}{\textit{10369.768}}                                         \\ \cline{1-2} \cline{4-5} 
\multicolumn{1}{|c|}{( 1,1,0 )}                                                          & \multicolumn{1}{c|}{7652.108}                                                   & \multicolumn{1}{l|}{} & \multicolumn{1}{c|}{( 1,1,0 )}                                                          & \multicolumn{1}{c|}{10371.676}                                                  \\ \cline{1-2} \cline{4-5} 
\multicolumn{1}{|c|}{\textbf{( 0,1,1 )}}                                                 & \multicolumn{1}{c|}{\textit{7294.347}}                                          & \multicolumn{1}{l|}{} & \multicolumn{1}{c|}{( 0,1,1 )}                                                          & \multicolumn{1}{c|}{10371.541}                                                  \\ \cline{1-2} \cline{4-5} 
\multicolumn{1}{|c|}{( 1,1,1 )}                                                          & \multicolumn{1}{c|}{8317.976}                                                   & \multicolumn{1}{l|}{} & \multicolumn{1}{c|}{( 1,1,1 )}                                                          & \multicolumn{1}{c|}{10373.750}                                                  \\ \cline{1-2} \cline{4-5} 
                                                                                         & \multicolumn{1}{l}{}                                                            &                       &                                                                                         & \multicolumn{1}{l}{}                                                            \\
                                                                                         & \multicolumn{1}{l}{}                                                            &                       &                                                                                         & \multicolumn{1}{l}{}                                                           
\end{tabular}
\end{table}

Table \ref{tab:aic_arima} shows that the best ARIMA model for ANF has $p=0$, $d=1$, and $q=1$ also known as \textit{simple exponential smoothing} (SES) model\footnote{https://people.duke.edu/~rnau/411arim.htm\#ses} which corrects autocorrelated errors. It is worth note that, for some nonstationary time series (e.g., ones that exhibit noisy fluctuations around a slowly-varying mean), the random walk model does not perform as well as a moving average of past values. In other words, rather than taking the most recent observation as the forecast of the next observation, it is better to use an average of the last few observations in order to filter out the noise and more accurately estimate the local mean. The SES model uses an exponentially weighted moving average of past values to achieve this effect. EOG has $p=0$, $d=1$, and $q=0$, also known as \textit{random walk}\footnote{https://people.duke.edu/~rnau/411arim.htm\#arima010} ( \cite{danyliv2019random}), where $y_{t+1}=y_t+\epsilon_t$, and $\epsilon_t$ are a sequence of centered, uncorrelated random variables. An ARIMA(0, 1, 0) series, when differenced once ($d=1$), becomes an ARMA(0, 0), which is random, uncorrelated, and noise. If $X_1,X_2,X_3,\dots$ are the random variables in the series, this means that $X_{i+1}-X_i=\epsilon_{i+1}$ where $\epsilon_1,\epsilon_2,\dots$ are a sequence of centered, uncorrelated random variables, so $X_{i+1}=X_i+\epsilon_i$ reveals that we have a \textit{Random Walk}.

\begin{figure}[ht]
	\centering
\begin{minipage}{.50\hsize}
  \includegraphics[width=1.\linewidth,left]{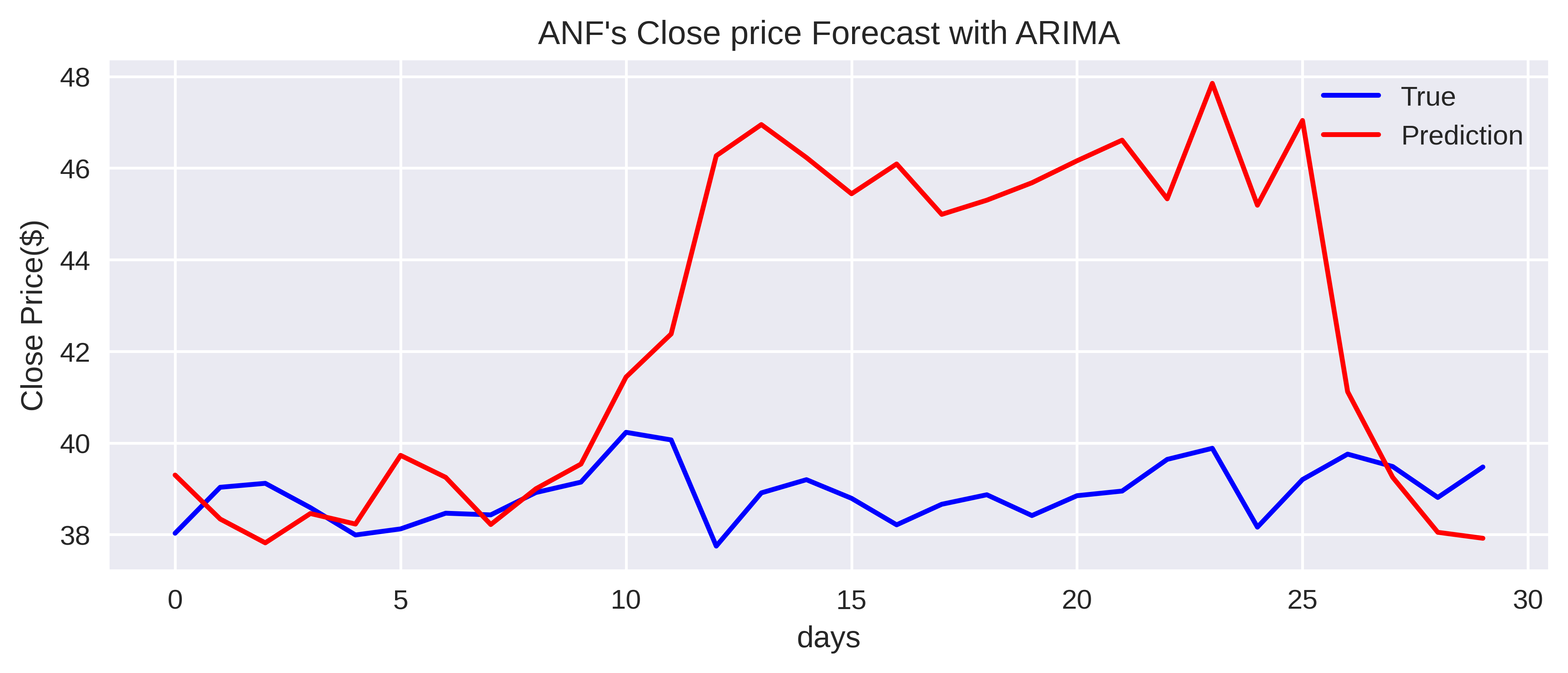}
\end{minipage}%
\begin{minipage}{.50\hsize}
  \includegraphics[width=1.\linewidth,right]{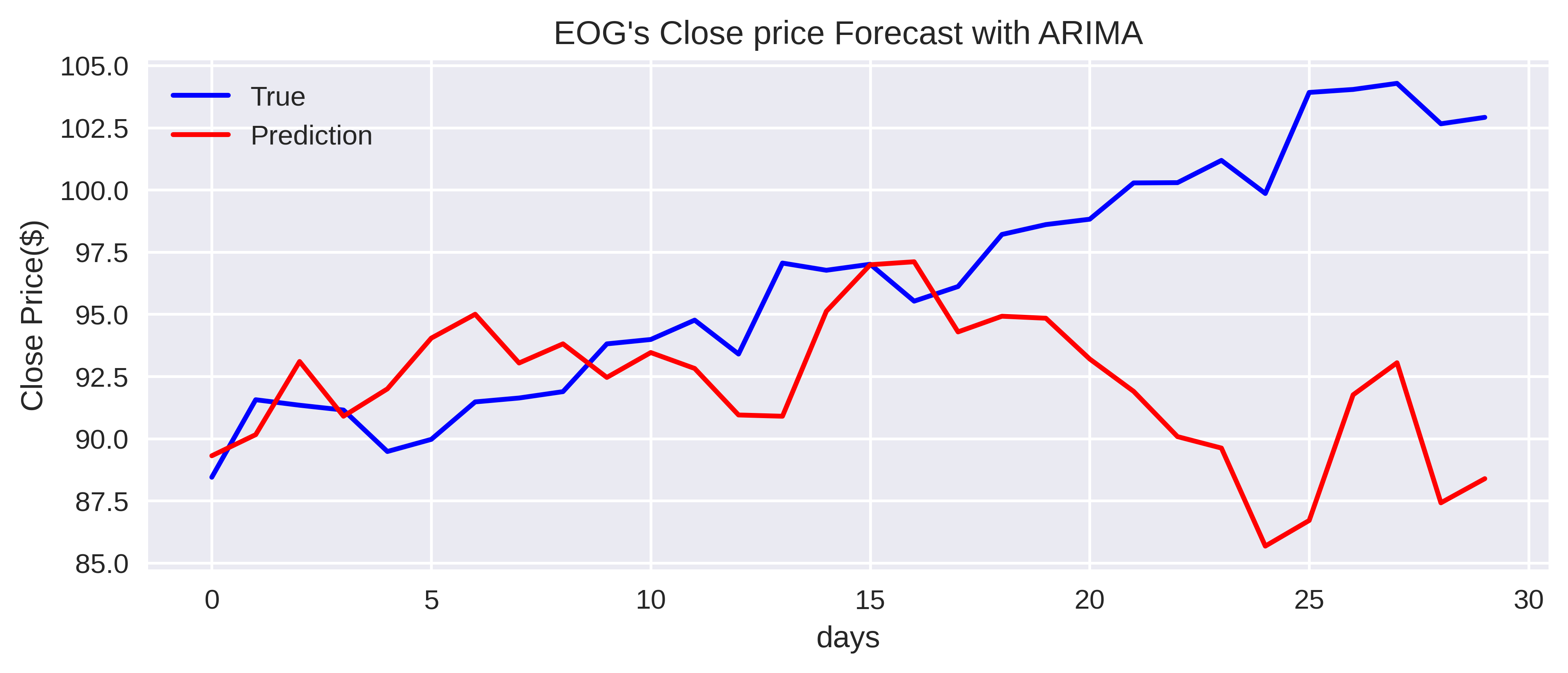}
\end{minipage}
\caption{Detail of ARIMA forecast for the last 30 days of the ANF (left) and EOG (right) stock closing price.}\label{fig:30daysForecast}
\end{figure}

Fig. \ref{fig:30daysForecast} shows the predictions on the closing prices made with such auto-selected optimal ARIMA model for $n=30$ days following the training timespan, which corresponds to the 2507 days of market from October 30, 2011 to October 16, 2021. Note that, in the following, for the sake of brevity we will omit the graphs and values of low, high, and open prices, since the forecast errors are always very similar between the four OLHC components. 

\begin{table}[ht]
\centering
\caption{Error metrics of ARIMA on ANF and EOG stock price prediction.}
\label{tab:error_arima}
\begin{tabular}{lccccc}
\multicolumn{6}{c}{\textbf{ARIMA}}\\\hline
  & \textbf{MSE} & \textbf{RMSE} & \textbf{MAE} & \textbf{MAPE} & \textbf{EVS}\\ \hline
\textit{ANF}         & 25.49    & 5.05      & 3.86         & 0.09         & -0.02\\ \hline
\textit{EOG}         & 56.23    & 7.50      & 5.42         & 0.06         & -3.94\\ \hline
\end{tabular}
\end{table}

Table \ref{tab:error_arima} reports very high error metrics. It is worth noting that the model performs a bit better with the EOG stock, but not enough to be considered as a suitable tool for forecasting. 


\subsection{Prophet model}
\label{sec:prophet}

Looking for a statistical method to improve the ARIMA's results, we turned to Facebook Prophet (see \cite{Prophet2020Facebook}). Prophet is "a procedure for forecasting time series based on an additive model where non-linear trends are fit with yearly, weekly, and daily seasonality, plus holiday effects. It works best with time series that have strong seasonal effects and several seasons of historical data. Prophet is robust to missing data and shifts in the trend, and typically handles outliers well." \cite{ProphetWeb}.


More technically, Prophet is an additive regressive model which uses a time series with three main
components: trend, seasonality, and holidays, combined in the equation $y(t)=g(t)+s(t)+h(t)+\epsilon(t)$, where 
$g(t)$ is the trend function which models non-periodic changes in the value of the
time series, $s(t)$ represents the seasonality, i.e., periodic changes (e.g., the number of trades might also depend on the month/year),
$h(t)$ represents the effects of holidays, which have a clear impact on most business time series, and $\epsilon(t)$ is the  error term, following a normal distribution.

In our experiments, we used Prophet "out of the box", leaving all the default parameter selections.

\begin{figure}[ht]
	\centerline{\includegraphics[width=1.0\hsize]{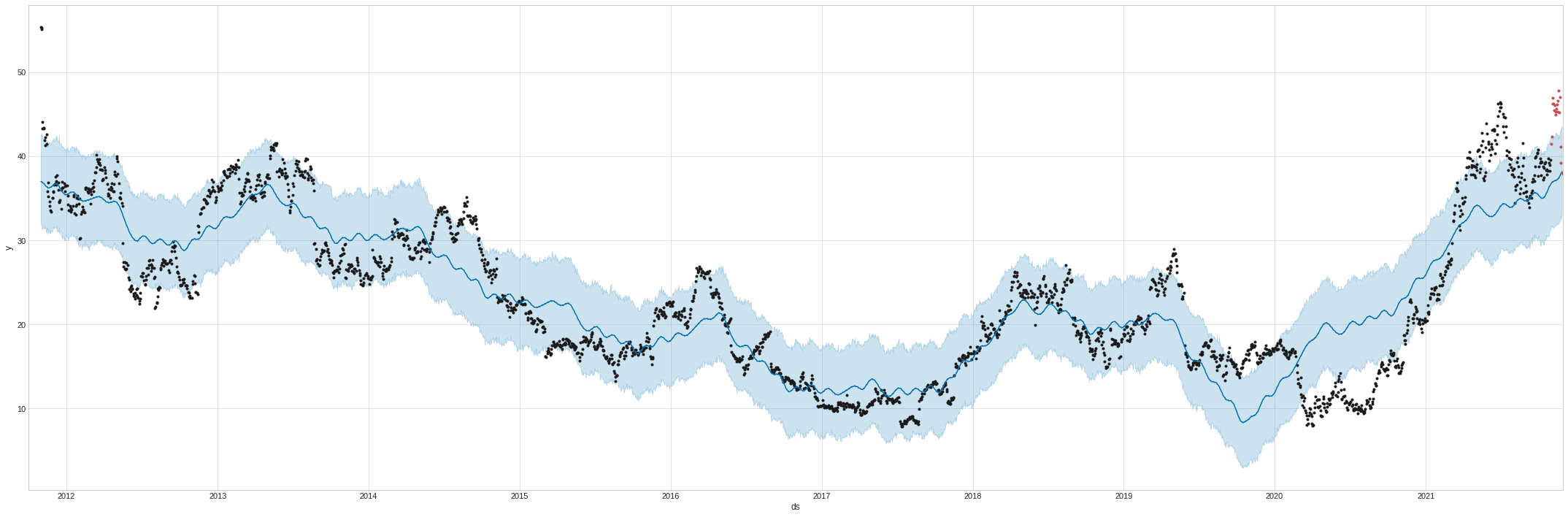}}
	\caption{Detail of Prophet forecast for the ANF stock closing price.}
	\label{fig:ANFprophet}
\end{figure}

\begin{figure}[ht]
	\centerline{\includegraphics[width=1.0\hsize]{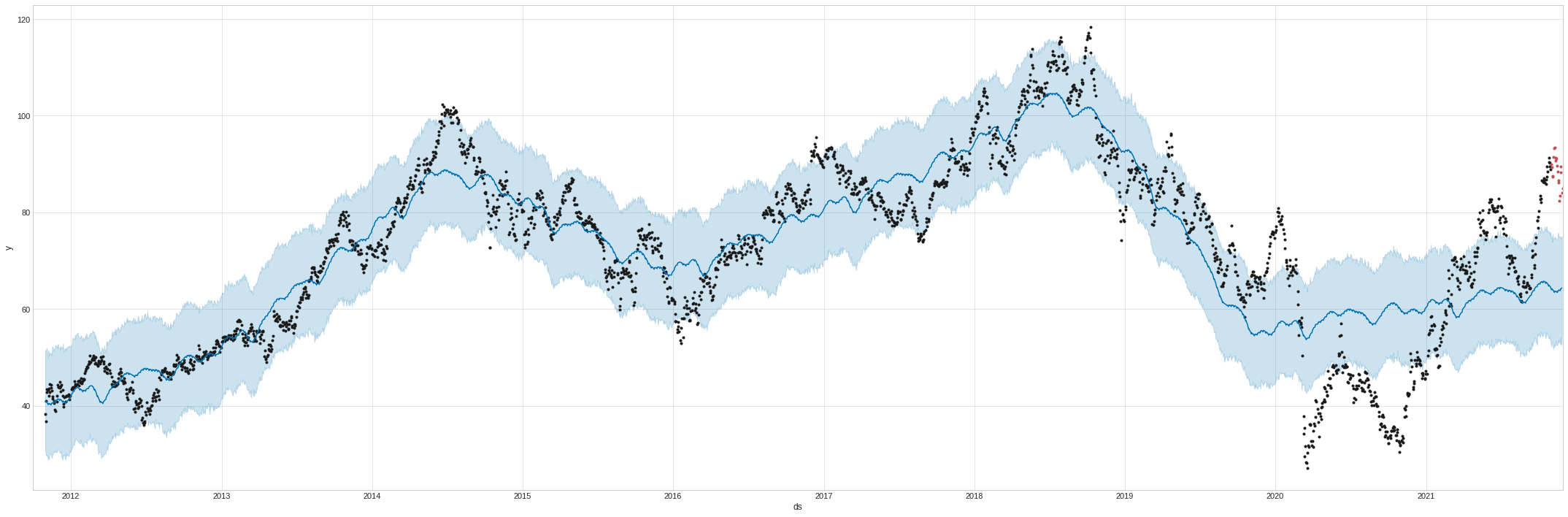}}
	\caption{Detail of Prophet forecast for the EOG stock closing price.}
	\label{fig:EOGprophet}
\end{figure}

Figs. \ref{fig:ANFprophet} and \ref{fig:EOGprophet} show for the ANF and EOG closing prices, respectively the historical data as black dots, the future actual data as the rightmost red dots, and the Prophet forecasts as a blue line.

\begin{figure}[ht]
	\centerline{\includegraphics[width=1.0\hsize]{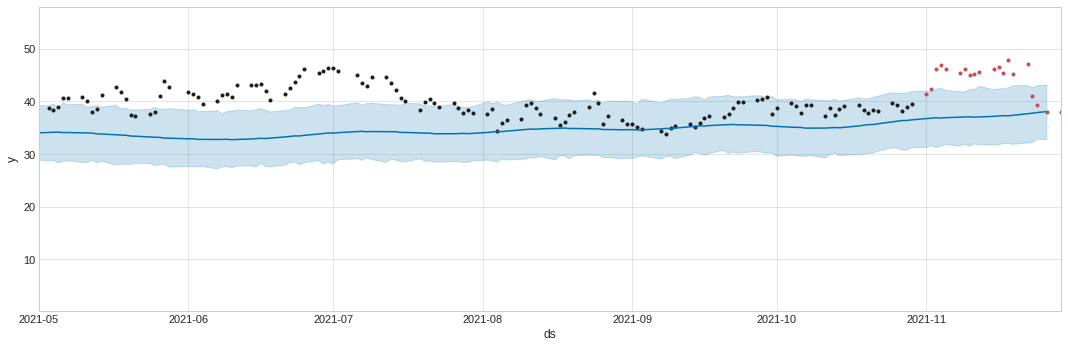}}
	\caption{Detail of the last five months of Prophet forecast for the ANF stock closing price.}
	\label{fig:ANFprophetDetail}
\end{figure}
\begin{figure}[ht]
	\centerline{\includegraphics[width=1.0\hsize]{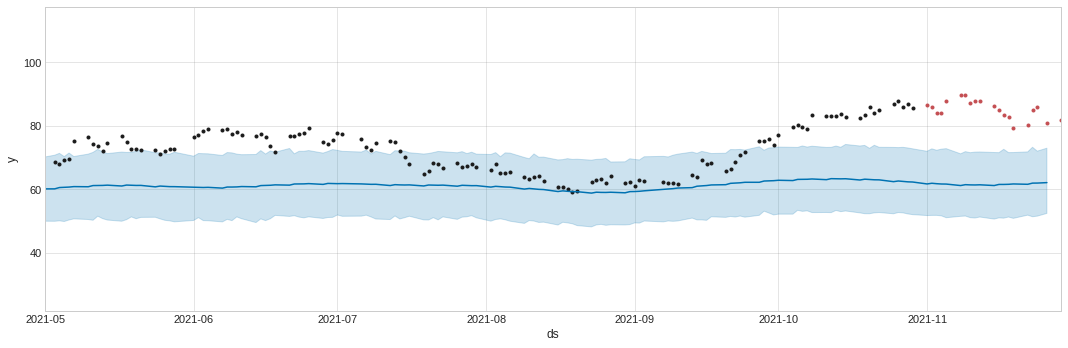}}
	\caption{Detail of the last five months of Prophet forecast for the EOG stock closing price.}
	\label{fig:EOGprophetDetail}
\end{figure}

It is clear that the model performs well in the first years but, when volatility increases (approximately from year 2020), the forecasts start to be clearly worse in the latest period as shown in detail in figs. \ref{fig:ANFprophetDetail} and \ref{fig:EOGprophetDetail}.

\begin{table}[ht]
	\centering
	\caption{Error metrics of Prophet on ANF and EOG stock price prediction.}
	\label{tab:error_prophet}
	\begin{tabular}{lccccc}
		\multicolumn{6}{c}{\textbf{Prophet}}\\\hline
		& \textbf{MSE} & \textbf{RMSE} & \textbf{MAE} & \textbf{MAPE} & \textbf{EVS}\\ \hline
		\textit{ANF}         & 53.04    & 7.28      & 6.71         & 0.16         & 0.31\\ \hline
		\textit{EOG}         & 713.61    & 26.71      & 26.54         & 0.30         & -0.03\\ \hline
	\end{tabular}
\end{table}

Table \ref{tab:error_prophet} reports the corresponding error metrics calculated in the same time frame used for the ARIMA experiments. Clearly, the performances are unacceptable also in this case.

\newpage

\section{Forecasting with Deep Learning Methods}
\label{sec:deeplearning}

Compared with conventional artificial neural networks, deep neural networks are characterized by a higher number of neurons and hidden layers each of which, in principle, gives the network a greater ability to extract high-level features.
This makes DNNs very efficient in solving nonlinear problems: in particular, when it comes to time series forecasting, DNNs can fill the gap left open by traditional statistical techniques such as the ones presented in Section \ref{sec:statistical}, which often assume that the series are generated by linear processes and consequently may be inappropriate for most real-world problems that are overwhelmingly nonlinear.
Indeed, works like \cite{yao1999neural} and \cite{hansen1999time}  (which focuses on time series prediction) show that neural network models often outperform conventional ARIMA models, and in particular \cite{hansen1999time} also show that neural networks outperform ARIMA in predicting the direction of stock movement, since are able to detect hidden patterns in the time series.  

\subsection{The Deep Neural Networks}
In this subsection, we maintain the same forecasting objective of Section \ref{sec:statistical}, i.e., $n=30$ days following the training date, grouped owever, in \textit{batches} of $bs=5$, corresponding to the working days in an open market week. We use four DNNs, one for each OHLC price, all with the same architecture. 
However, while the Auto-ARIMA detected that the optimal configuration for such algorithm was to generate a forecast based on the previous value (see Section \ref{sect:arima}, here we empirically found that the neural network performs better if its \textit{input layer} is fed with the $t=5$ previous values, i.e., the prices of the previous market week. In other words, to forecast the price of a day $s$, the input neurons will be presented to the prices of days $s-1, \ldots, s-5$, respectively.

When building a neural network for applications like financial forecasting, one must find a compromise between \textit{generalisation} and \textit{convergence}. For example, hidden layers must not have too many nodes, since they may lead the DNNs to learn the training data without performing any generalization. 
Therefore, to find the geometry (number and size of hidden layers) which minimizes the error on all the networks, we developed a python module that generates  different network geometries in combination with the sklearn GridSearchCV\footnote{https://scikit-learn.org/stable/modules/generated/sklearn.model\_selection.GridSearchCV.html} algorithm, which in turn tries to find the optimal combination of the hyperparameters (epochs, batch size, learning rate, optimizer employed) for each specific network.

The resulting optimal geometry has two hidden layers composed by $10*t$ and $5*t$ neurons respectively, as in \cite{LetteriPG18} and \cite{Letteri2019journal}. 
Finally, each network outputs its price prediction for the two considered stocks 
via a single neuron in the output layer.

To help reducing overfitting, we also applied a dropout of $0.2\%$ on each of the two internal layers (\cite{Geoffrey2012Hinton}). In addition, to introduce non-linearity between layers, we used \textit{ReLU} as the activation function, which performs better than a \textit{tanh} or \textit{sigmoid} functions (\cite{Krizhevsky2012Alex}), despite the fact that the depth of the network consists of only a few internal layers.

To estimate the network learning performance during the training we use the \textit{L1loss} function,  which measures the mean absolute error (MAE) between each predicted value and the corresponding real one. The optimization algorithm used to minimize such loss function during the training is the \textit{adaptive moment} (Adam), an extension of the \textit{stochastic gradient descent} (SGD).

The trained networks are freely available and can be downloaded from our repository\footnote{https://www.ivanletteri.it/dnnModels/}.

\begin{figure}[ht]
	\centering
	\begin{minipage}{.51\hsize}
		\includegraphics[width=1.\linewidth,left]{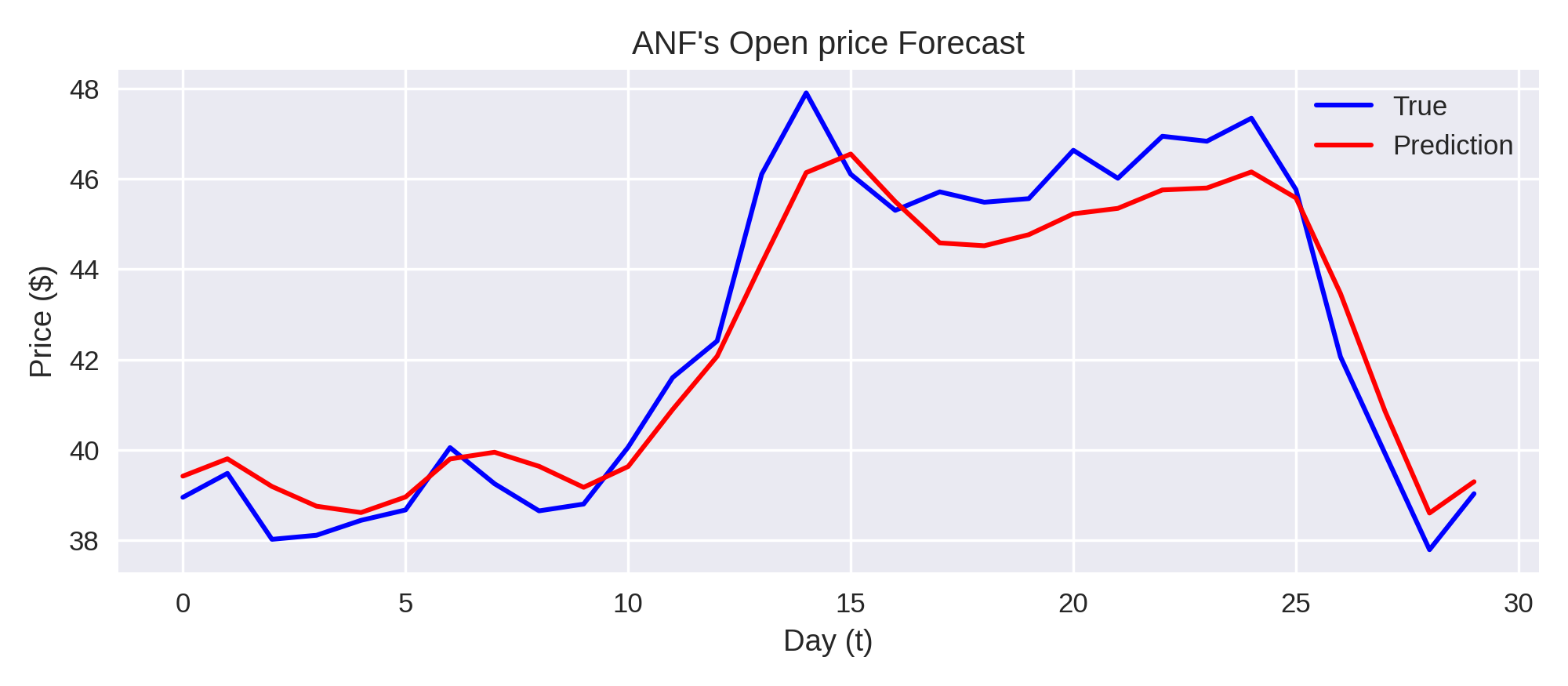}
	\end{minipage}%
	\begin{minipage}{.51\hsize}
		\includegraphics[width=1.\linewidth,right]{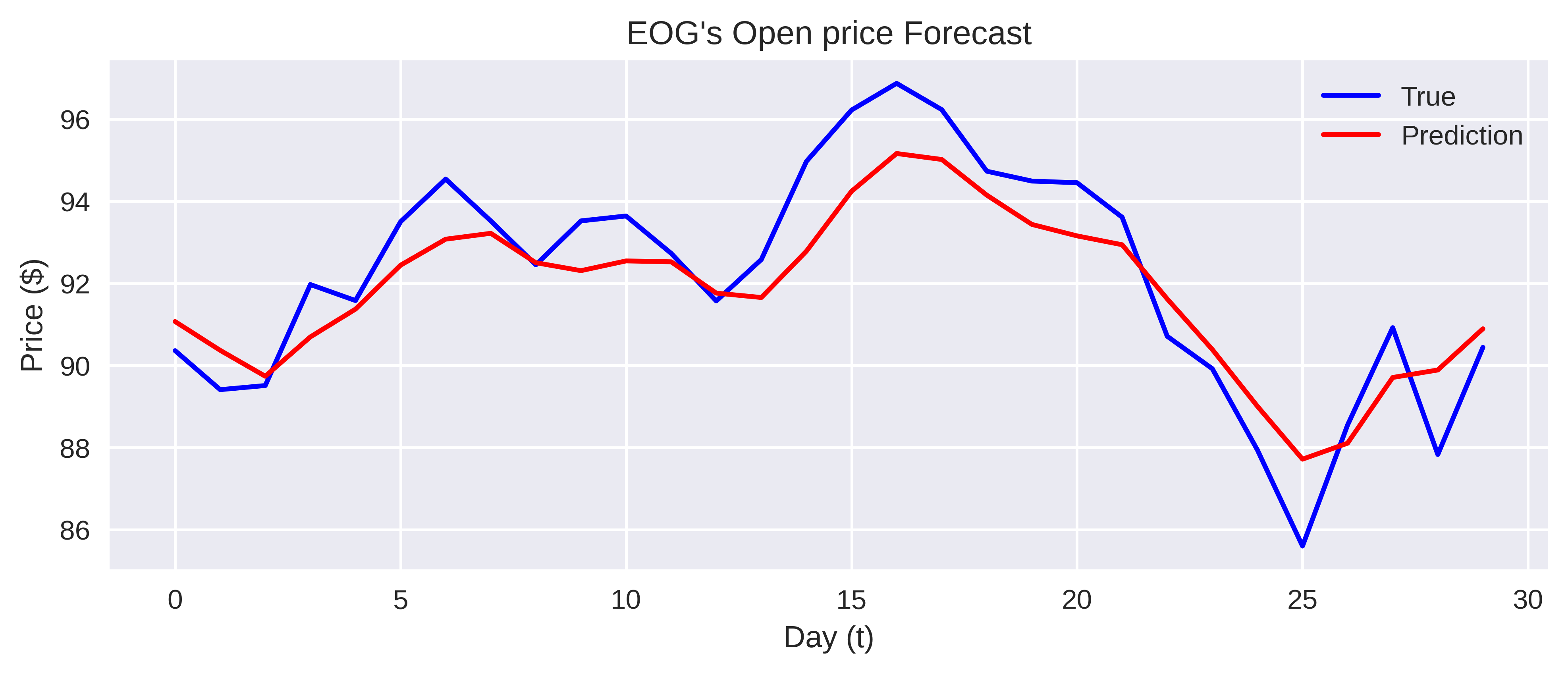}
	\end{minipage}
	\caption{DNNs prediction of 30 day of ANF and EOG open prices (validation set)}\label{fig:30daysopen}
\end{figure}

\begin{figure}[ht]
	\centering
	\begin{minipage}{.51\hsize}
		\includegraphics[width=1.\linewidth,left]{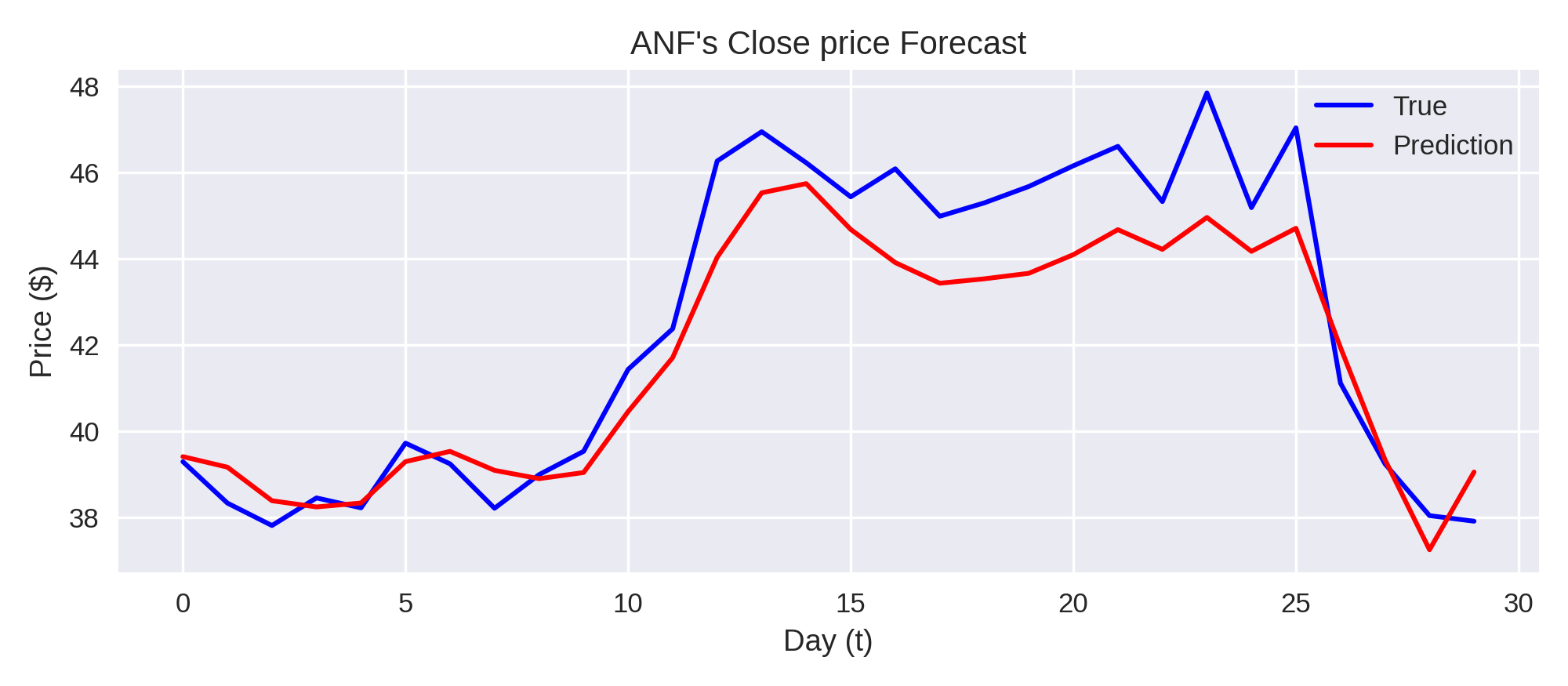}
	\end{minipage}%
	\begin{minipage}{.51\hsize}
		\includegraphics[width=1.\linewidth,right]{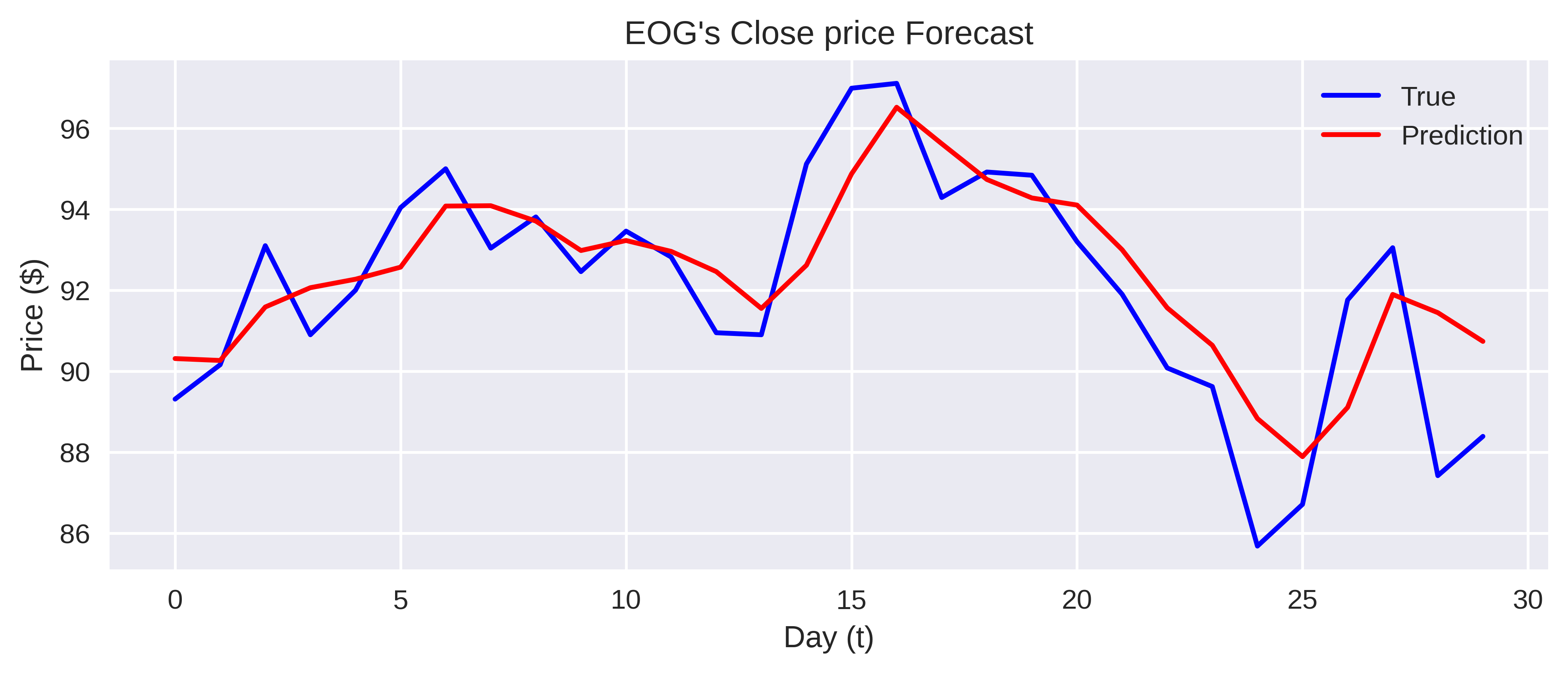}
	\end{minipage}
	\caption{DNNs prediction of 30 day of ANF and EOG close prices (validation set)}\label{fig:30daysclose}
\end{figure}

\begin{table}[ht]
\centering
\caption{Error metrics of DNN on ANF and EOG stock \textit{Close} price prediction (validation set)}
\label{tab:error_comparison}
\begin{tabular}{lccccc}
\multicolumn{6}{c}{\textbf{DNN}}\\\hline
  & \textbf{MSE} & \textbf{RMSE} & \textbf{MAE} & \textbf{MAPE} & \textbf{EVS}\\ \hline
\textit{ANF}         & 1.75    & 1.32      & 1.07         & 0.02         & 0.91\\ \hline
\textit{EOG}         & 2.39    & 1.55      & 1.23         & 0.01         & 0.7\\ \hline
\end{tabular}
\end{table}

\begin{figure}[ht]
	\centering
	\begin{minipage}{.51\hsize}
		\includegraphics[width=1.\linewidth,left]{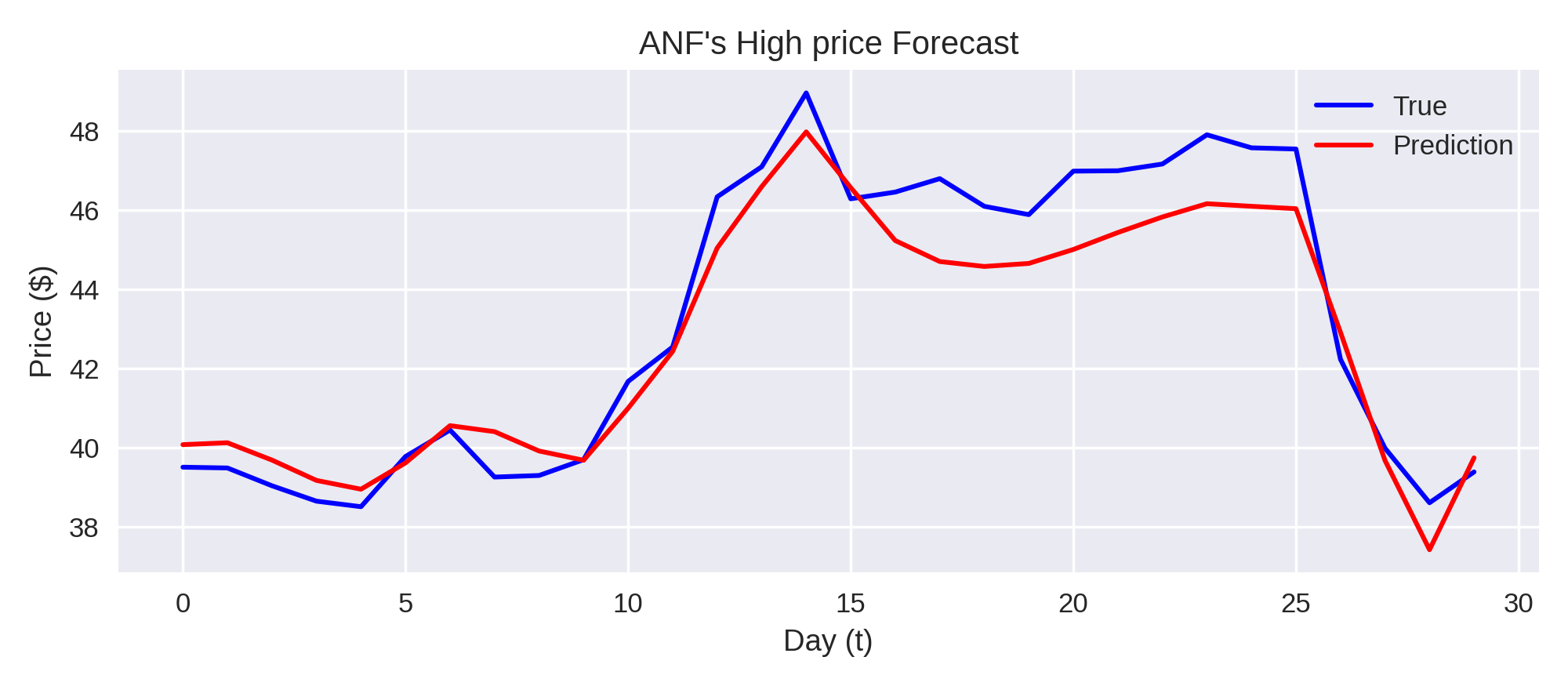}
	\end{minipage}%
	\begin{minipage}{.51\hsize}
		\includegraphics[width=1.\linewidth,right]{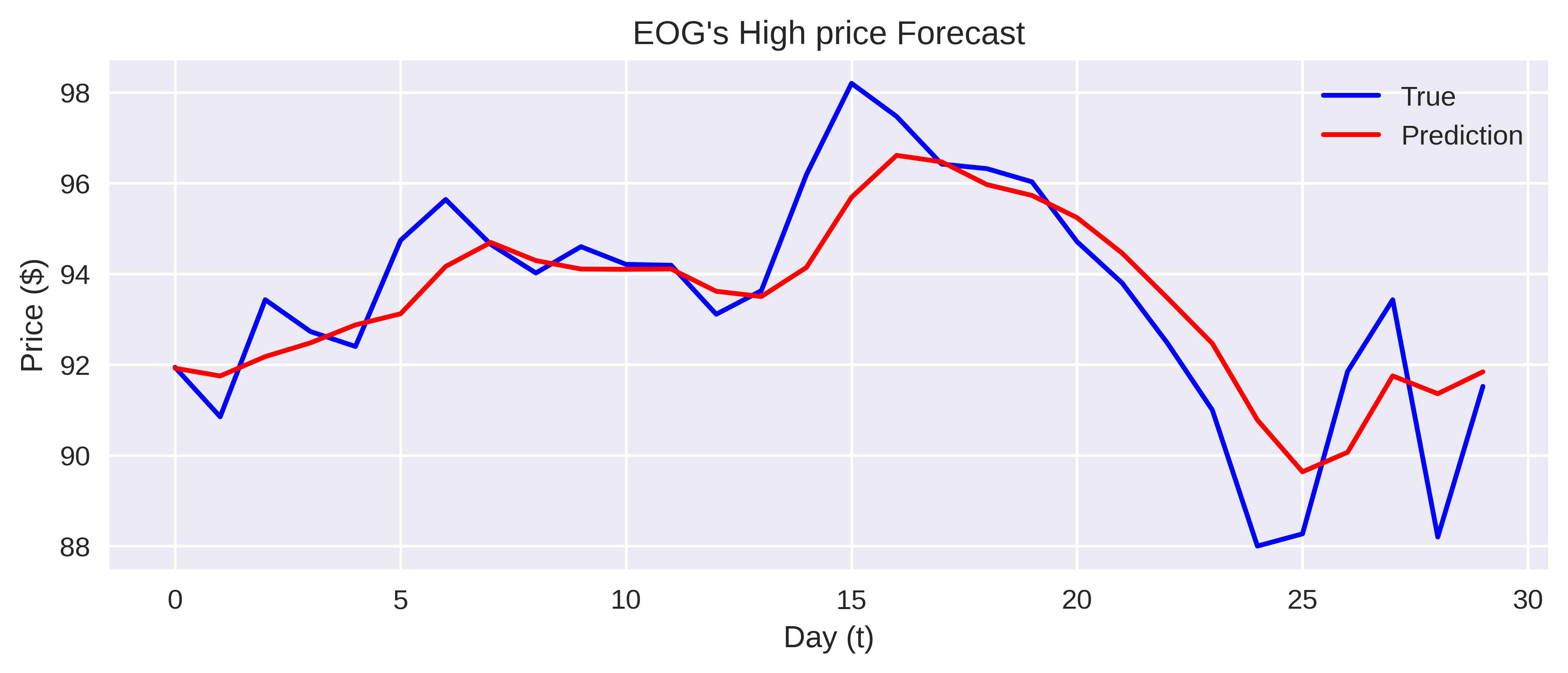}
	\end{minipage}
	\caption{DNNs prediction of 30 day of ANF and EOG high prices (validation set)}\label{fig:30dayshigh}
\end{figure}

\begin{figure}[ht]
	\centering
	\begin{minipage}{.51\hsize}
		\includegraphics[width=1.\linewidth,left]{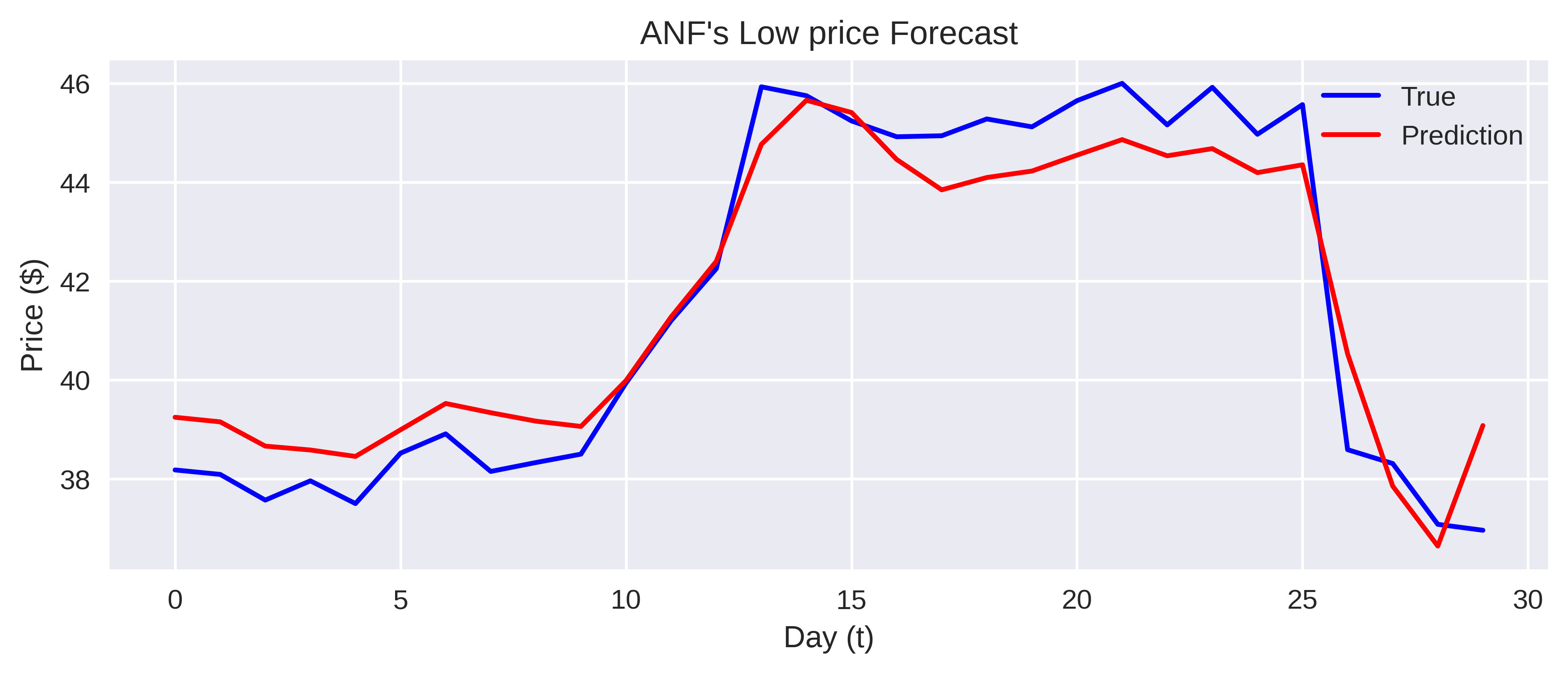}
	\end{minipage}%
	\begin{minipage}{.51\hsize}
		\includegraphics[width=1.\linewidth,right]{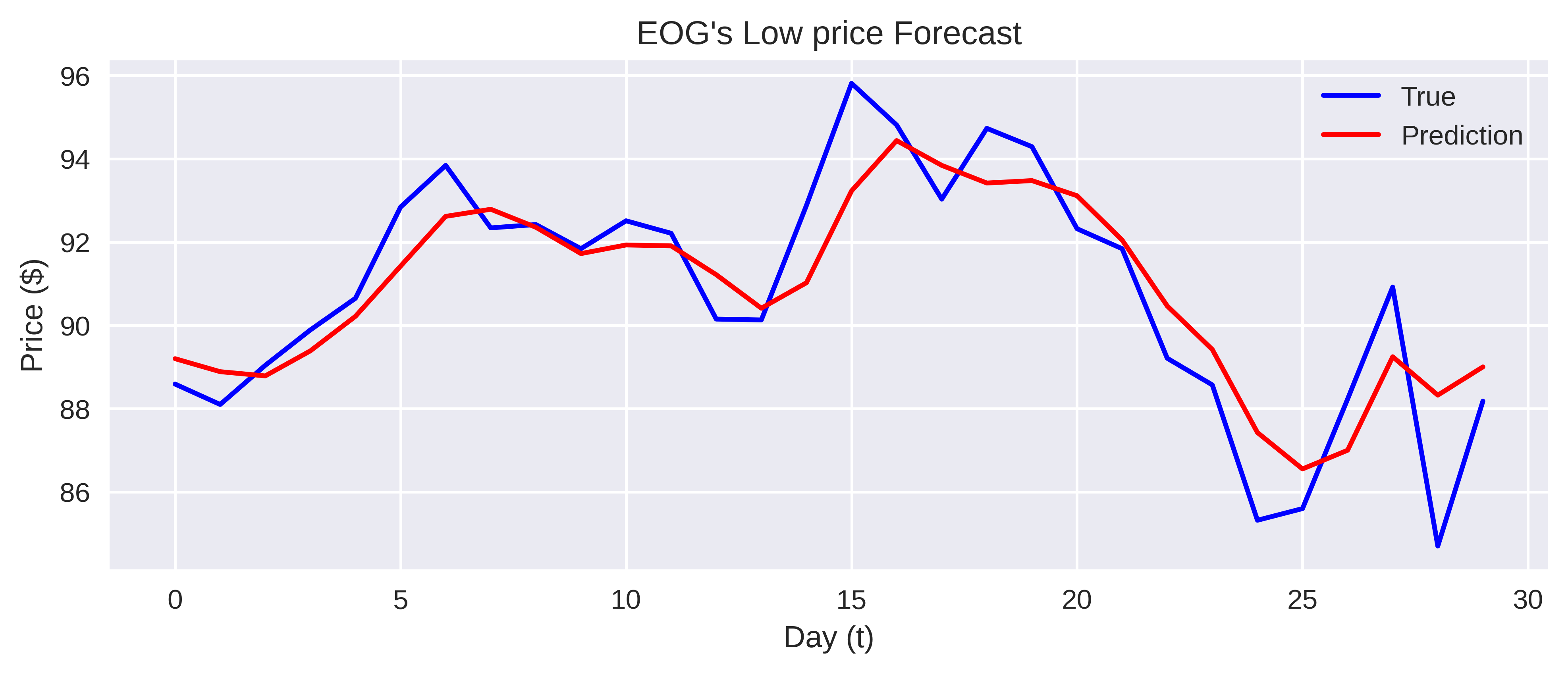}
	\end{minipage}
	\caption{DNNs prediction of 30 day of ANF and EOG low prices (validation set)}\label{fig:30dayslow}
\end{figure}

Figs. \ref{fig:30daysopen}, \ref{fig:30dayshigh}, \ref{fig:30dayslow}, and \ref{fig:30daysclose} show the DNN forecasts on the ANF and EOG open, high, low, and close prices, respectively, in the same 30-day time frame used for the experiments of the previous section. For the sake of brevity, we report only the corresponding error metrics about close prices in tab. \ref{tab:error_comparison}.
It is clear that the DNN performs better than the statistical methods shown in the previous section.

\subsection{DNN-Forwardtesting Experiment}

After showing that DNNs are the best forecast model for our stock price dataset, we want to show the performances of a trading system that is based on such forecasts. 
A trading \textit{strategy} tells the investor when to buy or sell shares in such a way that the sequence of these operations is profitable. 
Typically, discretionary traders base such strategy on the values of one or more technical indicators. System traders, which use algorithms to guide their trading, typically apply a rule-based approach, where rules are also based on a set of technical indicators.
In both cases, such indicators are usually chosen by the trader using the so called \textit{backtesting} technique, i.e., by considering the available historical data and choosing the indicator(s) so that the corresponding strategy would get the highest profit if applied on the past.

On the other hand, we propose a novel, alternative approach, which exploits the DNN forecasts and selects the indicators such that the corresponding trading strategy would get the highest profit on the \textit{possible future} given by the forecasts. Our hypothesis is that the DNN forecasts may encode a deeper understanding of the past trends, i.e., we actually exploit the historical data in a way that the traditional approach would not be able to do. 

In particular, the (algorithmic) trading strategy of our system is encoded in a set of  \textit{entry} and \textit{exit} \textit{trading rules} which are in turn based on the value of a single technical indicator chosen from a set of ten common technical indicators, i.e., Simple Moving Average (SMA), Exponential Moving Average (EMA), Moving Average Convergence Divergence (MACD), Bollinger Bands (BBs), Stochastic (ST), William \%R (W\%R), Momentum (MO),  Relative Strength Index (RSI), Average true range (ATR), Price Oscillator (PO) (see \cite{Barnwal2019Avinash}), Triple Exponential Moving Average (TEMA, \cite{Tsantekidis2017Avraam})  and Average Directional Index (ADX). We also tested some further meaningful combinations of the above indicators, like in \cite{prasad2022optimal}, and \cite{Hryshko2004Downs}, such as ST+MO+MACD, PO+W\%R, PO+RSI.

\subsubsection{Results} 

The best indicator for ANF is the  \textit{Triple Exponential Moving Average}, whereas \textit{Average Directional Index} is more suitable for EOG. In particular, 
\begin {itemize}
\item the Triple Exponential Moving Average (TEMA) (\cite{Tsantekidis2017Avraam}) is generally used to make short-term and medium-term intraday trading decisions to enter long or short trades, depending on bullish or bearish signals.  It was designed to smooth price fluctuations, making it easier to identify trends without the lag associated with moving averages.


\item the Average Directional Index (ADX)\footnote{https://en.wikipedia.org/wiki/Average\_directional\_movement\_index} is used  to determine the strength of a trend, and is accompanied by two further indicators: the negative directional indicator (-DI) and the positive directional indicator (+DI). As an example, when the +DI crosses above the -DI line, the rate of positive price change in the market is greater than the negative price change. If this happens when the ADX is above 25, it is a solid signal to place buy orders. Similarly, when the -DI crosses above the +DI line, it implies that the rate of negative price change in the market is greater than the positive price change. If this happens when the ADX is below 25, it is a solid signal to place sell orders. 
\end{itemize}

\begin{figure}[ht]
\centering
\small
\begin{tabular}{lll}
\hline
\textbf{ANF}\\\cline{2-3}
&\textit{Entry} & $((x^{(l)} < TEMA^{(l)}) \vee (x^{(h)} < TEMA^{(h)})) \wedge ((x^{(c)} < TEMA^{(c)}) \vee (x^{(o)} < TEMA^{(o)}))$ \\\cline{2-3}
&\textit{Exit} & $((x^{(l)} > TEMA^{(l)}) \vee (x^{(h)} > TEMA^{(h)})) \wedge ((x^{(c)} > TEMA^{(c)}) \vee (x^{(o)} > TEMA^{(o)}))$ \\\hline
\textbf{EOG}\\\cline{2-3}
&\textit{Entry} & $(+DI > -DI) \wedge (ADX > 25)$\\\cline{2-3}
&\textit{Exit} & $(-DI > +DI) \wedge (ADX > 25)$\\\hline
\end{tabular}
\caption{Trading system rules.}
\label{fig:trading}
\end{figure}

The final trading rules, based on such indicators, are shown in fig. \ref{fig:trading}, where $(o)$, $(h)$, $(l)$,  $(c)$ refer to the OHLC prices, respectively, and $x$ is the current (opening, highest, etc.) price.

\begin{table}[!ht]
\centering
\caption{Performance of trading system on validation set with ANF and EOG stocks.}
\label{tab:results}
\begin{tabular}{lcccccc}
\hline
& \textbf{\begin{tabular}[c]{@{}c@{}}Num. of\\ Trades\end{tabular}} & \textbf{\begin{tabular}[c]{@{}c@{}}Total\\ Return (\$)\end{tabular}} & \textbf{\begin{tabular}[c]{@{}c@{}}Expectancy\\ Ratio \end{tabular}} & \textbf{\begin{tabular}[c]{@{}c@{}}Sharpe\\ Ratio\end{tabular}} & \textbf{\begin{tabular}[c]{@{}c@{}}Sortino\\ Ratio\end{tabular}} & \textbf{\begin{tabular}[c]{@{}c@{}}Calmar\\ Ratio\end{tabular}} \\ \hline
ANF    & 3         & 6.126           & 2.112                       & 2.194              & 3.340                        & 12.403               \\ \hline
EOG    & 3         & 1.374           & 0.525                       & 1.253              & 2.556                        & 5.814              \\ \hline
\end{tabular}
\end{table}

Then, we evaluated the profit derived from the application of such a strategy on the \textit{real} data of the 30-day trading period following October 16, 2021, having as starting point a budget of \$100 reinvested in compounded mode. The results are shown in Table \ref{tab:results}.

\subsubsection{Baseline comparison}
To compare our strategy with a baseline, we re-evaluated the same set of  technical indicators through the traditional backtesting technique on the historical data for the 30 days \textit{before} October 16, 2021, to see if it would result in different choices and maybe different profits. 

\begin{figure}[!ht]
\centering
\begin{minipage}{.5\textwidth}
  \centering
  \includegraphics[width=1.\linewidth]{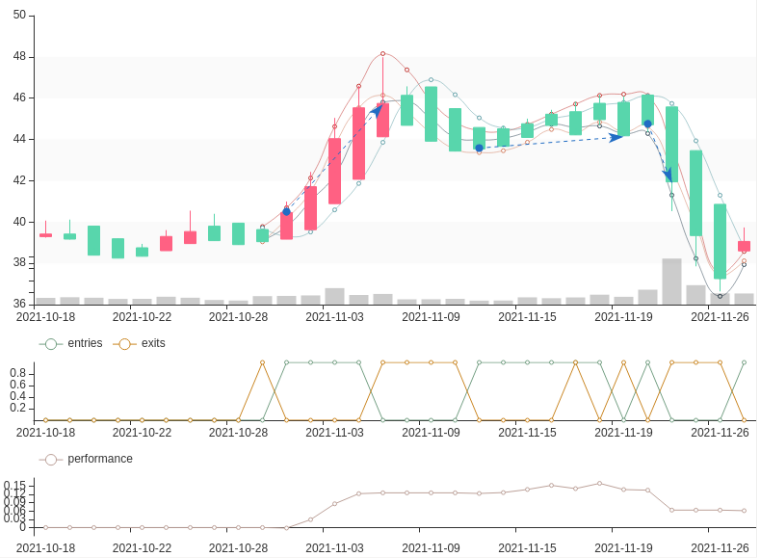}
  \label{fig:ANFtrades_predicted}
\end{minipage}%
\begin{minipage}{.5\textwidth}
  \centering
  \includegraphics[width=1.\linewidth]{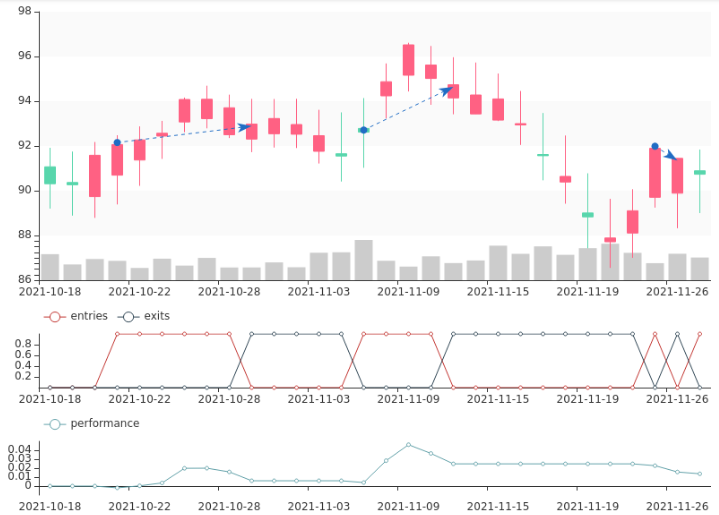}
  \label{fig:EOGtrades_predicted}
\end{minipage}
\caption{Performance results on predicted ANF OHLC values to the left and EOG OHLC values to the right (N.B.:
the candlestick graphs follow the asian notation).}
\label{fig:trades}
\end{figure}
\begin{figure}[ht]
	\centering
	\begin{minipage}{.5\hsize}
		\includegraphics[width=1.\linewidth,left]{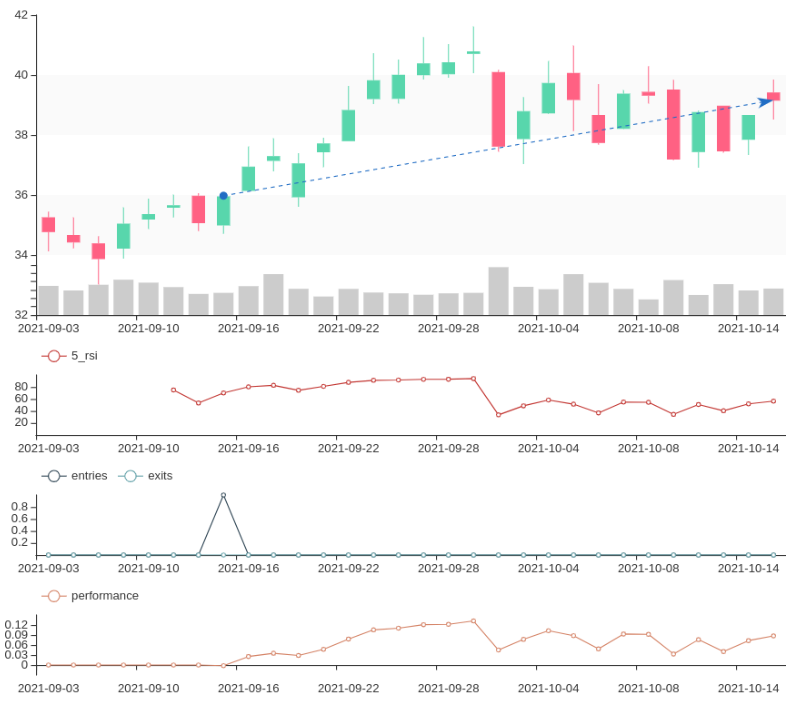}
	\end{minipage}%
	\begin{minipage}{.5\hsize}
		\includegraphics[width=1.\linewidth,right]{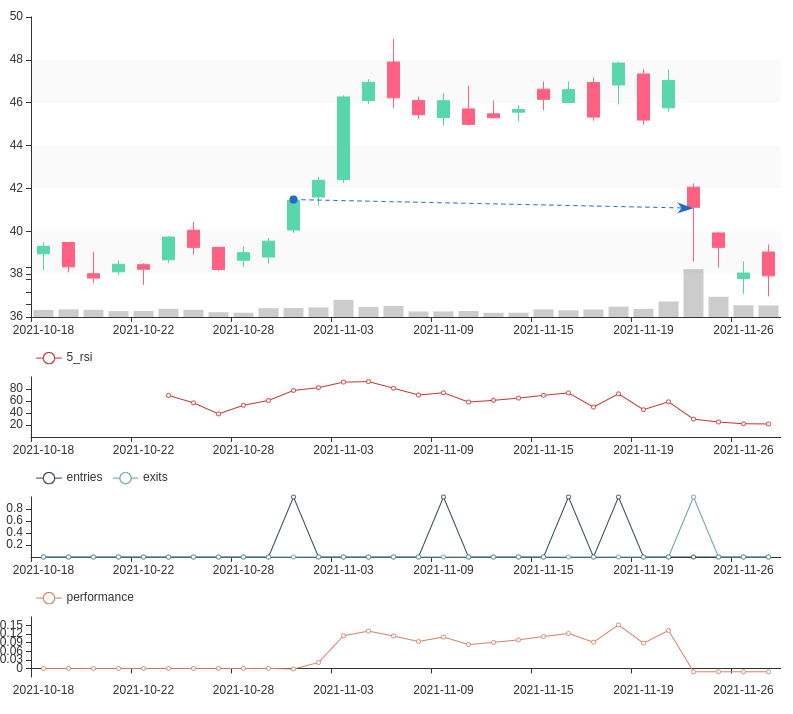}
	\end{minipage}
	\caption[RSI indicator on ANF close prices last 30 days of trainset (to the left) and 30 days in the future (to the right)]{RSI indicator on ANF close prices last 30 days of trainset (to the left) and 30 days in the future (to the right).}\label{fig:ANF_rsiPriceClose}
\end{figure}

The details plotted using library Finlab\footnote{https://pypi.org/project/finlab-crypto/} in fig. \ref{fig:trades}, shown that a trader using backtesting would choose ADX for the EOG share, as with DNN-forwardtesting technique, so the profit would be the same in this case. However, the TEMA indicator would not be chosen for the ANF share. Indeed, the most promising indicator, given the past 30 days of market, would be RSI with time period $5$, overbought=70, oversell=30 as in fig. \ref{fig:ANF_rsiPriceClose} but, if applied to the future, it would result in a lower profit.

\begin{table}[!ht]
\centering
\caption{RSI profit observing the latest 30 days price of ANF stocks.}
\label{tab:ANFbestTI}
\begin{tabular}{lcccccc}
\hline
& \textbf{\begin{tabular}[c]{@{}c@{}}Num. of\\ Trades\end{tabular}} & \textbf{\begin{tabular}[c]{@{}c@{}}Total\\ Return (\$)\end{tabular}} & \textbf{\begin{tabular}[c]{@{}c@{}}Expectancy\\ Ratio \end{tabular}} & \textbf{\begin{tabular}[c]{@{}c@{}}Sharpe\\ Ratio\end{tabular}} & \textbf{\begin{tabular}[c]{@{}c@{}}Sortino\\ Ratio\end{tabular}} & \textbf{\begin{tabular}[c]{@{}c@{}}Calmar\\ Ratio\end{tabular}} \\ \hline
ANF 30 days before   & 1         & 8.741           & 1.421                       & 2.266              & 3.064                        & 20.238               \\ \hline
ANF 30 days after   & 1         & -1.168           & -1.1683                       & 0.119              & 0.158                        & -0.935               \\ \hline
EOG 30 days before   & 1         & 8.741           & 1.421                       & 2.266              & 3.064                        & 20.238               \\ \hline
EOG 30 days after   & 1         & -1.168           & -1.1683                       & 0.119              & 0.158                        & -0.935               \\ \hline
\end{tabular}
\end{table}


\section{Conclusion}
In this paper, we propose a stock market trading system that exploits deep neural networks as part of its main components improving a previous work (\cite{letteri2022stock}). 

In such a system, the trades are guided by the values of a pre-selected technical indicator, as usual in algorithmic trading. However, the novelty of the presented approach is in the indicator selection technique: traders usually make such a selection by \textit{backtesting} the system on the historical market data and  choosing the most profitable indicator with respect to the \textit{known past}. On the other hand, in our approach, such most profitable indicator is chosen by \textit{DNN-forwardtesting} it on the \textit{probable future} predicted by a deep neural network trained on the historical data. 

As discussed in the paper, neural networks outperform the most common statistical methods in stock price prediction: indeed, their predicted future allows to make a very accurate selection of the indicator to apply, which takes into account trends that would be very difficult to capture through backtesting. 

To validate this claim, we applied our methodology on two very different assets with medium volatility, and the results show that our DNN forwardtesting-based trading system achieves a profit that is equal or higher than the one of a traditional backtesting-based trading system.

Given the promising potentials of this approach, we will further test its reliability on other stock markets using different data, such as cryptocurrencies or defi-tokens, also varying the timeframes for day trading and scalping activities.

In the near future we plan to explore data analysis following a proven process and feature selection strategy (see \cite{LetteriPVG20}, \cite{LetteriPC19}). The new datasets will certainly require a balancing (e.g., buy, sell and hold trades). This balancing will take place according to the algorithm called \textit{Generative 1 Nearest Neighbours} (\cite{LetteriArxivG1No}) and targeted oversampling (\cite{LetteriCDP21}).

In the future work, we consider to analyze DNNs considering them some black-box decision-making systems, which inherently introduce the risk of impacting aspects such as ethics (\cite{DyoubCLL21}) further when spread to Multi Agent System (\cite{DyoubICLP21}), and we will analyze them from a Logic-based Machine Learning perspective (\cite{Dyoub2020Letteri}).

\bibliographystyle{unsrtnat}
\bibliography{references}  

\end{document}